\documentclass[letterpaper,11pt]{report}

\usepackage{fullpage}
\usepackage{verbatim}
\usepackage{cite}
\usepackage{setspace}
\usepackage{fancyhdr}

\usepackage[small]{caption}
\usepackage{graphics}
\usepackage{color}

\usepackage[hyphenbreaks]{breakurl}
\usepackage[hyphens]{url}

\usepackage[dvips]{graphicx}
\usepackage{amssymb}
\usepackage{epstopdf}
\usepackage{subfigure}
\usepackage{multirow}
\usepackage{booktabs}
\usepackage{longtable}

\def\title{Call Me MayBe: Understanding Nature and Risks of Sharing Mobile Numbers on Online Social Networks}

\begin{document}

\def\addrone{Your address}
\def\addrtwo{Your city}

\def\degree{M.Tech. in Computer Science with Specialization in Information Security}

\def\submissiondate{January 08, 2011}

\def\supervisorone{Dr Ponnurangam Kumaraguru}

\def\supervisortwo{Dr Amarjeet Singh}

\def\supervisorthree{Dr Alessandra Sala}

\thispagestyle{empty}

\begin{center}

{\LARGE \bf {Call Me MayBe: \\
Understanding Nature and Risks of Sharing Mobile Numbers on Online Social Networks}

 }
 \vspace{.3in}

 {\Large{Prachi Jain}} \\
 \vspace{.1in}
 IIIT-D-MTech-CS-IS-11-007 \\

 Nov 14, 2013 \\

    \vspace{.35in}

  \vspace{.25in}

{Indraprastha Institute of Information Technology\\
New Delhi}

\vspace{.35in}  {\underline{Thesis Committee} \\ \supervisorone                ~(Chair) \\ \supervisortwo \\ \supervisorthree} \\ \vspace{.35in}
\vspace{1in}

 {Submitted in partial fulfillment of the requirements \\for the Degree of M.Tech. in Computer Science, \\ with specialization in Information Security}

\vspace{.2in}

\copyright 2013 by IIIT Delhi \\ All rights reserved \\
\vspace{.8in}

\end{center}

This research was partially funded by Precog Group at IIIT Delhi.

\newpage

\pagestyle{empty}
\vspace*{7.1in}
General Terms: Measurement, Security, Design, Human Factor
Keywords: Online Social Networks, Privacy, Mobile Number, Phone number, Risks

\newpage

\begin{center}
\section*{Certificate}\label{section:certificate}
\end{center}
This is to certify that the thesis titled \textbf{``Call Me MayBe: Understanding Nature and Risks of Sharing Mobile Numbers on Online Social Networks"} submitted by \textbf{Prachi Jain} for the partial fulfillment of the requirements for the degree of \emph{Master of Technology} in \emph{Computer Science \& Engineering} is a record of the bonafide work carried out by her under my guidance and supervision in the Security and Privacy group at Indraprastha Institute of Information Technology, Delhi. This work has not been submitted anywhere else for the reward of any other degree. \\ \vspace{0.5in}

\textbf{Dr Ponnurangam Kumaraguru}\\
\textbf{Indraprastha Institute of Information Technology, New Delhi}

\begin{abstract}
There is a great concern about the potential for people to leak private information on social networks.  There are many anecdotal examples of this, but few quantitative studies. This research explores the activity of sharing mobile numbers on OSNs, in particular via public posts. In this work, we understand the characteristics and risks of mobile numbers sharing behaviour on OSNs either via profile or public posts and focus on \emph{Indian} mobile numbers. We collected 76,347 unique mobile numbers posted by 85,905 users on Twitter and Facebook and analyzed 2,997 numbers, prefixed with +91. We observed that most users shared their own mobile numbers to spread urgent information; and to market products, IT facilities and escort business. Fewer females users shared mobile numbers on Online Social Networks. Users utilized other social networking platforms and third party applications like Twitterfeed and TweetDeck, to post mobile numbers on multiple OSNs. In contradiction to the user's perception of numbers spreading quickly on OSN, we observed that except for emergency, most numbers did not diffuse deep.

\indent To assess risks associated with mobile numbers exposed on OSNs, we used numbers to gain sensitive information about their owners (e.g. name, Voter ID) by collating publicly available data from OSNs, Truecaller, Open government data repository (OCEAN). On using the numbers on WhatApp, we obtained a myriad of sensitive details (relationship status, BBM pins, travel plans) of the mobile number owner. We communicated the observed risks to the owners by calling them on their mobile number. Few users were surprised to know about the online presence of their number, while few users intentionally posted it online for business purposes.~\footnote{Call Me MayBe: Understanding Nature and Risks of Sharing Mobile Numbers on Online Social Networks, Conference on Online Social Networks (COSN) 2013} We observed that 38.3\% of users who were unaware of the online presence of their number have posted their number themselves on the social network. With these observations, we highlight that there is a need to monitor leakage of mobile numbers via profile and public posts. To the best of our knowledge, this is the first exploratory study to critically investigate the exposure of Indian mobile numbers on OSNs.

\end{abstract}

\newpage
\pagestyle{empty}

\newpage
\noindent%
\begin{minipage}[c][\textheight][c]{\textwidth}%
\centering{Dedicated to my indefatigable parents Mr. Prakash Chand Jain and Mrs. Cheena Jain $\&$ my beloved siblings Nikita and Shubham.}%
\end{minipage}%

\newpage

\section*{Acknowledgments}\label{section:acknowledgments}
\pagestyle{plain}
\pagenumbering{roman}
I would like to express my deepest gratitude to Dr. Ponnurangam Kumaraguru (PK) for providing invaluable guidance and encouragement which enabled me to accomplish the work presented in this thesis. I would also like to thank him for providing significant discussions and valuable feedback on different aspects of my work. The flexible and helpful nature of my advisor made it easier for me to pursue this research work in a simple and exciting manner. I want to thank Dr. Alessandra Sala and Dr. Amarjeet Singh for agreeing to serve in my committee.\\
\\
I take this opportunity to thank Siddhartha Asthana and Precog Group at IIIT Delhi for their cooperation, discussions and suggestions to the work. A special  thanks to Paridhi Jain, a PhD student at IIIT Delhi for her involvement in the research discussions related to the work. I would also like to thank Anupama Aggarwal for patiently shepherding this report.\\
\\
Finally it is the support, encouragement and good wishes of my family without whom I would not have been able to complete my thesis. I thank and owe my deepest regards to all of them and all others who have helped me directly or indirectly.

\newpage

\tableofcontents
\listoffigures
\listoftables

\newpage

\newpage

\newpage
\mbox{}


\chapter{Introduction}\label{chapter:introduction}
\pagenumbering{arabic}
\setcounter{page}{1}
\onehalfspacing
Today, Online Social Networks (OSNs) have facilitated their users with variety of services. Users can easily connect to new people and re-connect to old friends, receive live feeds of their friends' activity, and share multimedia content in controlled and restrictive ways. These services have attracted users to spend substantial time (27\% of their online time) on OSNs~\cite{Timespent}. Users generate voluminous new content on OSNs, for instance, 46\% of adult Internet users post original photos or videos that they themselves have created~\cite{Pew}. User Generated Content (UGC) on online social networks is observed to have high similarities with offline interactions of users~\cite{Rowe:2010:CDI:1772938.1772947, subrahmanyam2008online}. Therefore, concerns have been raised on (un)intentional mention of one's sensitive information such as age, sexual orientation, travel patterns, credit card details, health records on online profile or posts~\cite{Mao:2011:LTA:2046556.2046558, Wang:2011:IRM:2078827.2078841,DBLP:conf/infocom/DeyTRS12,jain2013call}.

\begin{figure}[!ht] 
   \centering
   \subfigure[User posts his own mobile number on Twitter, along with auxiliary information such as bank name and customer number.]{
   \includegraphics[scale=0.30]{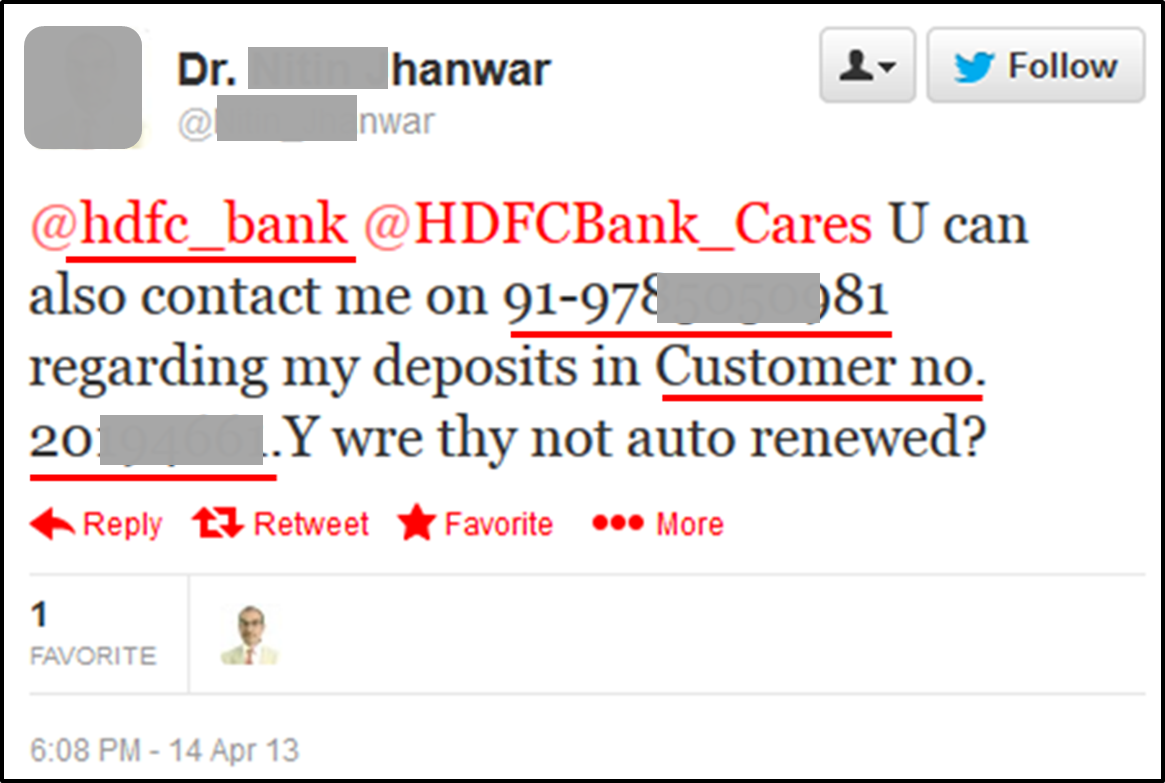}\label{fig:example1}
   }
   \quad
      \subfigure[Exposure of mobile number to unintended audience.]{
   \includegraphics[scale=0.30]{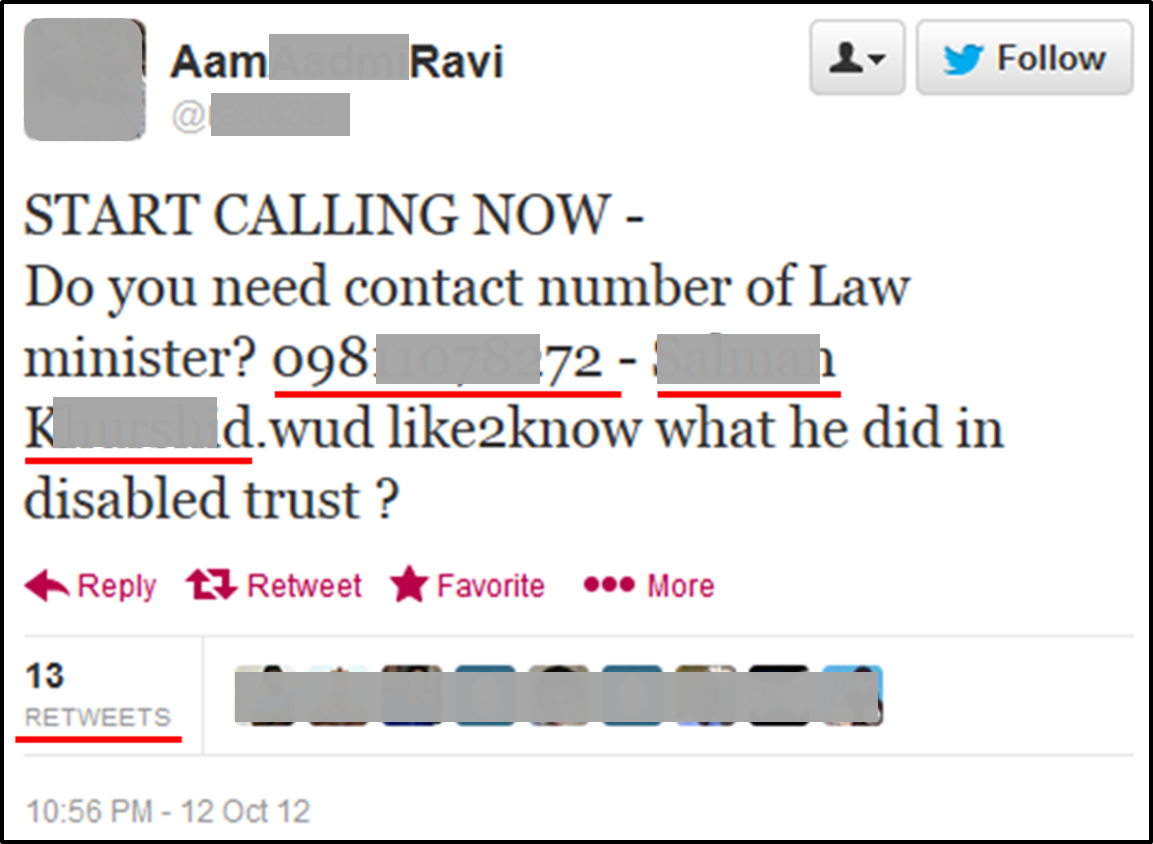}\label{fig:example3}
}
\centering
\quad
   \subfigure[User expose some other person's mobile number on Facebook, along with auxiliary information such as her name, her school's name and the city where she lives.]{
   \includegraphics[scale=0.30]{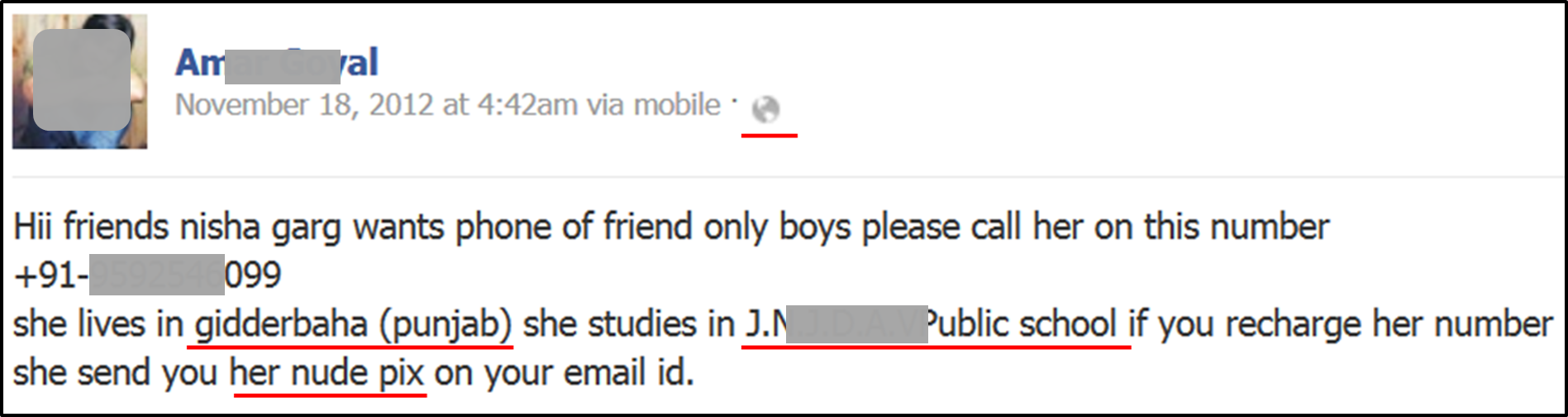}\label{fig:example2}
   }
   \caption{Public exposure of mobile numbers, along with axillary information}
   \label{fig:expose}
\end{figure}

Phone (Mobile) number is an example of identifiable information with which a real-world individual can be associated uniquely, in most cases~\cite{zheleva2011privacy,jain2013call}. The associated individual can become an easy target for SMS and phone-based phishing scams~\cite{Phishing}, spam SMSs~\cite{SMSspam}, spam calls~\cite{canada}, which may lead to annoyance, disturbance, stalking and denial of service. Such attacks can be made impactful with easy access to large number of mobile numbers shared publicly on OSNs. Mobile numbers can be shared either via profile attributes~\cite{Chen} or via posts (see Figure~\ref{fig:example1}). Auxiliary details of mobile number owners shared along with the mobile numbers, or collected otherwise, can help attackers to launch targeted attacks against them~\cite{Grandparentscam}. Figure~\ref{fig:example1} shows an example where a user posted his \emph{own} mobile number on Twitter, while complaining to his \emph{bank}, and revealed his \emph{customer number} too. Figure~\ref{fig:example2} shows an example where a user posted some \emph{other girl's} mobile number in an attempt to deface her, he also posted the \emph{girl's name}, her \emph{school's name} and the \emph{city} where she lived. Direct exposure of mobile numbers to unintended audience also amplifies the risk. Figure~\ref{fig:example3} shows an example where the number posted by the user is re-tweeted and becomes available to \emph{unintended audience}.

To examine the necessity of safeguard methods, which can be used to prevent public exposure of users' mobile numbers either via profile or posts, in this work we make an attempt to comprehend mobile number sharing behavior on OSNs, and the gravity of associated risks.

\indent India has been a popular venue for mobile and phone frauds~\cite{india2, Sim}, owing to huge telecom industry. India has the second largest mobile network with 919.17 million subscribers by Feb 2013.~\footnote{\url{http://www.trai.gov.in/trai/upload/PressReleases/566 /pr25apr08no43.pdf}} We therefore, focus on exposure of Indian mobile numbers in this work. We explore ~\emph{why}, ~\emph{how} and ~\emph{whereabouts} of Indian mobile numbers shared on two most popular OSNs -- Facebook and Twitter~\cite{PopularOSN}. An Indian mobile number can be used to reveal critical information about its owner such as name, age, location, which may invite targeted identity attacks. We communicate the perceived risks of sharing mobile numbers online to their owners by calling them on their mobile numbers and recording their reaction during the call.

\section{Organization}
\indent The report is organized as follows: Chapter~\ref{chapter:related_work} presents related work on privacy leaks on OSNs, and highlights research gaps. Chapter~\ref{chapter:Overview} enumerates contributions of the report, Chapter~\ref{chapter:methodology} describes the methodology we followed to collect mobile numbers shared on social networks. Chapter~\ref{chapter:analysis} elaborates the analysis and characteristics of leaked mobile number on OSNs. Chapter~\ref{chapter:augmentation} discusses if mobile numbers can be exploited to disclose other sensitive information of their owners. Chapter~\ref{chapter:discussion} presents quick summary of the report and a discussion on the applicability of the results 
and presents future directions, and some limitations of our work.

\chapter{Related Work}\label{chapter:related_work}
\onehalfspacing

To position this work, here is a brief overview of related work on understanding disclosure of identifiable information behaviour, risks associated with it and possible countermeasures of such disclosures.

~\section{Identity information and its disclosure on online social networks}
\indent On an online social network, a user is defined by a set of attributes e.g. name, age, education, and friendship network. A user's attributes can be classified into three sub categories -- identifying, quasi-identifying and sensitive attributes~\cite{zheleva2011privacy}. Identifying attributes connect an online user account to a unique real-world entity, for instance, email address. Quasi-identifying attributes are a set of attributes which when combined together with each other, connect to a unique real-world entity. Gender, zip code, and birthdate are quasi-identifying attributes, which together deanonymized 87\% of Americans~\cite{Sweeney:2002:KAM:774544.774552}. Sensitive attributes are user's characteristics which she intends to hide and does not wish to make them public, for example, health records, sexual orientation, current location.
In this report we refer to identifying attributes or quasi-identifying attributes as Personally Identifiable Information (PII). It is important to note that PII is an ever expanding category. Ten years ago, identifiers like Social Security Numbers were considered as PII but now literature has provided evidence of other features like movie rating being a potential PII~\cite{ohm2010broken}.

\indent Researchers have widely studied leakage of PII and sensitive attributes on OSNs e.g. email address~\cite{Balduzzi}, age~\cite{DBLP:conf/infocom/DeyTRS12}, gender~\cite{Burger:2011:DGT:2145432.2145568}, travel patterns~\cite{Wang:2011:IRM:2078827.2078841}, phone numbers~\cite{Magno:2012:NKB:2398776.2398794}, and group memberships~\cite{zheleva:www09}. Magno \textit{et al.} in their work on characterization of Google+ social network, observed many users shared mobile numbers as their profile attribute. They observed that single Indian males shared most mobile numbers~\cite{Magno:2012:NKB:2398776.2398794}. In our work, we attempt to dive deeper to understand the other characteristics of exposed Indian mobile numbers on OSNs.

~\section{Consequences of identity information disclosure on online social networks}

\indent Researchers have also discussed the possible outcomes of leakage of PII and sensitive attributes. Jagatic et al in their work demonstrated how publicaly available personal information revealed on online social networks, can be exploited for ~\emph{social} phishing ~\cite{jagatic2007social}. Besides this, PII and sensitive attribute leaks may further support identity disclosure attacks~\cite{Wondracek:2010:PAD:1849417.1849976}, linkage attacks~\cite{Chen}, and privacy attacks~\cite{Mao:2011:LTA:2046556.2046558,Wang:2011:IRM:2078827.2078841}. Disclosed friendship relations of a user can be used to deploy an automated identity theft attack ~\cite{bilge2009all}. Particularly information like mobile numbers can solely be used to exploit smartphone messaging services (which uses mobile numbers for authentication e.g. WhatsApp), and hence execute \emph{impersonation attack}, \emph{SMS spam attack}, \emph{Phone number enumeration attack} and \emph{Status message forgery attack}~\cite{schrittwieser2012guess}. VoIP applications like Viber~\footnote{Viber. \url{http://www.viber.com/}}, Voypi~\footnote{Voypi. \url{http://voypi.com/}}, Forfone ~\footnote{Forfone. \url{http://www.forfone.com/}}, EasyTalk, Wowtalk~\footnote{Wowtalk. \url{http://www.wowtalk.org/}} are also vulnerable to such attacks, when an attacker just have a mobile number in hand~\cite{schrittwieser2012guess}.

\indent On the flip side, mobile numbers themselves have been used to integrate all accounts on different smartphone messaging services like WeChat~\footnote{Wechat. \url{http://weixin.qq.com}} and MiTalk~\footnote{Mitalk messenger. \url{http://www.miliao.com/}} associated to a number, and bringing out more comprehensive information about mobile number users in China ~\cite{cheng2013bind}. However in this study we try to understand the comprehensiveness of information one can extract from a popular smartphone messaging service - WhatsApp using Indian mobile numbers leaked from Online Social Media.


\indent Apart from risks associated with attribute leakage, possible risks associated with aggregation of PII and sensitive attributes from multiple social networks, have been explored in the literature~\cite{ Chen,Mislove:2010:YYK:1718487.1718519, Irani:2009:LOS:1632709.1633521}. Krishnamurthy pointed in his work that auxiliary information collected from online sources could help in connecting an online profile uniquely to an offline entity~\cite{DBLP:journals/ieeesp/Krishnamurthy13}. In this work, we intend to explore the viability of the opinion. We exploit Indian mobile numbers posted on OSNs and attempt to understand if mobile numbers can be used to gather wider profile (e.g. name, location, age) of their owners.\\
\vspace{-2mm}
~\section{Communicating the risk of identity information disclosure}

\indent Krishnamurthy suggested that data augmentation privacy leaks could be prevented via alerting users about dispersive information sharing vulnerabilities. We follow the suggestion and attempt to communicate risks of online sharing of mobile numbers to their owners. Researchers have attempted to send Short Message Service (SMS) to mobile number owners~\cite{Avivore}, or publicly display the BBM pins or anonymized mobile numbers with the online profiles to embarrass the users~\cite{weKnow,bbm}. 
In this work, we communicated the risks by calling a sample of users (2,492) whose mobile numbers are available on social networks, via an Interactive Voice Response System (IVR). We chose IVR to ensure quick reachability of the risk communication message to the number owners. We captured every reaction of the users, when informed about the risks associated with sharing mobile numbers on OSNs and analyzed their responses.

\chapter{Contributions}\label{chapter:Overview}
\onehalfspacing
To understand the nature of mobile number sharing phenomenon on Online Social Networks and its associated risks, we do an intense analytical study. The major observations and contributions of the study are:
\begin{enumerate}
\item We observed Emergency, Marketing, Entertainment and Escort business were major contexts on Twitter however the context of marketing of IT facilities was observed on Facebook, when a user shared mobile number on OSNs.
\item We found that 
      users shared their own mobile numbers on OSNs more often than sharing other person's mobile number; and users of metropolitan cities in India actively posted mobile numbers on OSNs than other locations in India.
\item We showed that users post mobile numbers on multiple OSNs simultaneously, evident by the use of other social networks and third party applications to post content on Twitter and Facebook. We also observed that mobile numbers diffused deeper in Twitter when shared in emergency context than in other contexts.
\item We practically demonstrated the capability of an attacker to gain more comprehensive information about a user by using his mobile number as a starting point. We showed how leaked mobile numbers can be exploited to expose sensitive details of their owners such as number, age, voter ID, family details, complete address, and online activity.
\item Our experiments show that address book resolution feature, adopted by many smart phone messaging/VoIP applications today, can be exploited to link mobile numbers leaked from OSN profile with the corresponding profile on the messaging application and gain a wider profile of the mobile number owner. The experiment also showed how one can track when was the mobile number owner last online, hence a privacy breach for him.
\item We proposed a systematic approach for ~\emph{Risk Communication}. The risks of sharing of mobile numbers online, was communicated to their owners by calling them using an IVR system. We observed that 107 users were unaware of the online presence of their number, while, few were aware and told us that they posted the number intentionally for business purposes.
\item We found that out of 107 users who did not know that their number can be leaked, 38.3\% have posted their number themselves on the OSNs.
\end{enumerate}

\chapter{Methodology}\label{chapter:methodology}
\begin{figure*}[!ht] 
   \centering
   \includegraphics[scale=.65]{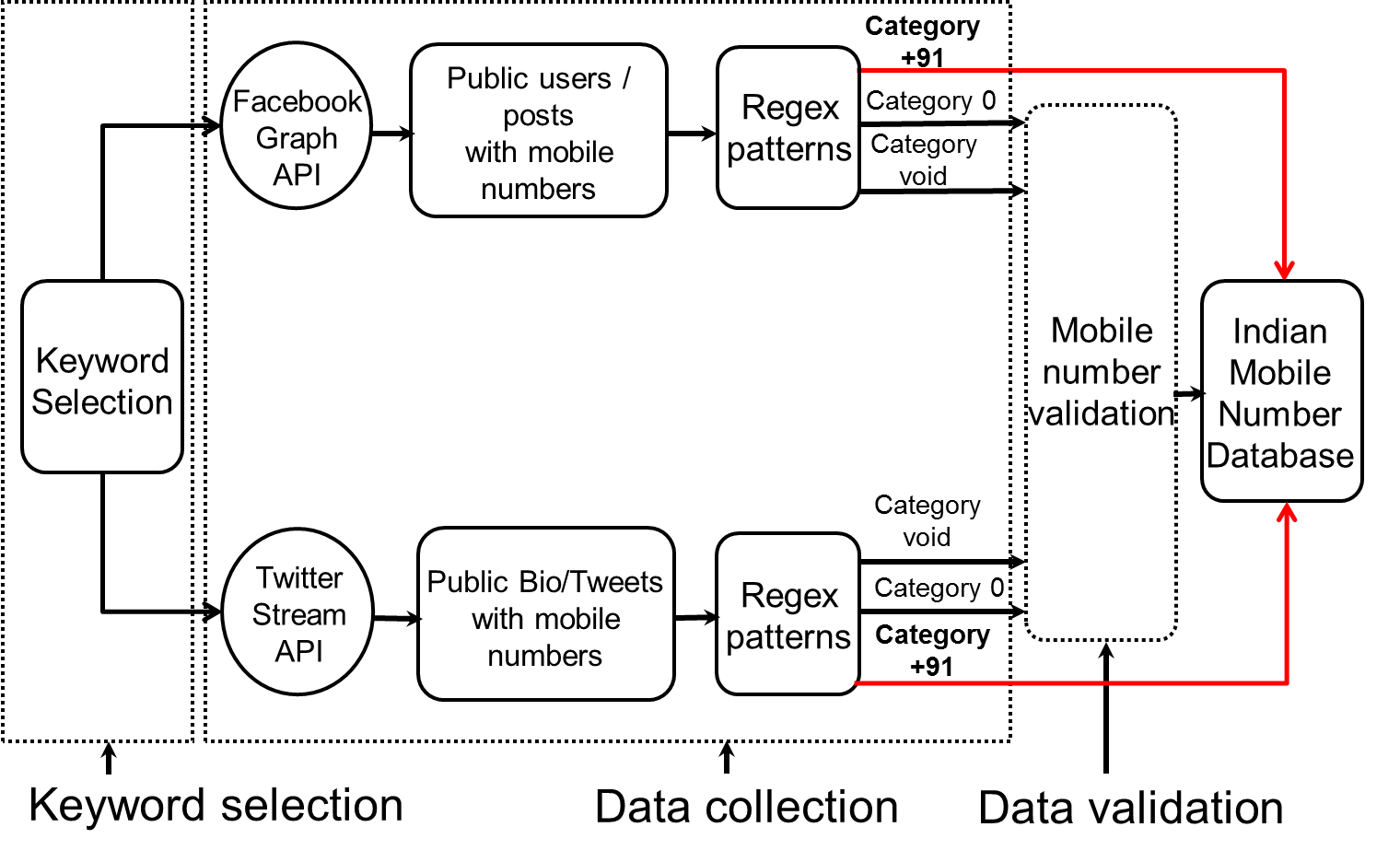}
   \caption{Data collection methodology to gather public profiles and posts which shared Indian mobile numbers on OSNs.}
   \label{fig:architecture}
\end{figure*}

\onehalfspacing
\indent We deployed an intelligent three stage socio-computational system to collect data suitable for analysis. The three stages of the system are -- keyword selection, data collection and data validation (see Figure~\ref{fig:architecture}). We collected Indian mobile numbers shared through posts or user description on two major and popular online social networks in India and across the globe -- Facebook and Twitter.

\section{Keyword selection}
\indent A pre-requisite to collect and filter public posts and tweets with a mobile number, was to select a set of relevant keywords. To create the keyword list, we surveyed a set of OSN users in IIIT-Delhi~\footnote{\url{http://iiitd.ac.in/}} to determine possible keywords they would use while sharing a mobile number on OSNs. We selected most commonly listed words for our initial set of 50 keywords such as \textit{mobile number}, \emph{contact us}, \emph{call me}. With the initial set of keywords, we collected 1,525 public tweets using Twitter Streaming API~\footnote{\url{https://dev.twitter.com}} and 1,000 public posts using Facebook Graph API.~\footnote{\url{https://developers.facebook.com/docs/reference/api}} We used the collected posts to identify other common keywords present when mobile numbers were shared (adapting a standard technique of query expansion from Information Retrieval \cite{Xu:1996:QEU:243199.243202}). We tokenized the posts, removed stop words and added most frequent words to expand the seed keyword set size to 278. Complete set of keywords are listed in Appendix (Table~\ref{table:Keyword_twitter}). Similar approach was used by Mao \textit{et al.} to gather tweets with required contexts~\cite{Mao:2011:LTA:2046556.2046558}.

\section{Data collection}
\indent We used the final set of keywords to collect public English posts and bio~\footnote{\raggedright{referred to as ``description" in Twitter API \url{https://dev.twitter.com/docs/platform-objects/users}}} which shared mobile numbers, using 
Twitter Streaming API and Facebook Graph API. We started our data collection from Facebook on November 16, 2012 and ended on April 20, 2013, while we started our data collection from Twitter on October 12, 2012 and ended on April 20, 2013.  We stored bio descriptions and public posts which shared mobile numbers on OSNs, along with profiles of the users who shared the number either via bio or public post.

\begin{figure*}[!htbp] 
    \centering
   \subfigure[Pattern 1: No space / dash in mobile number]{
   \includegraphics[scale=0.3]{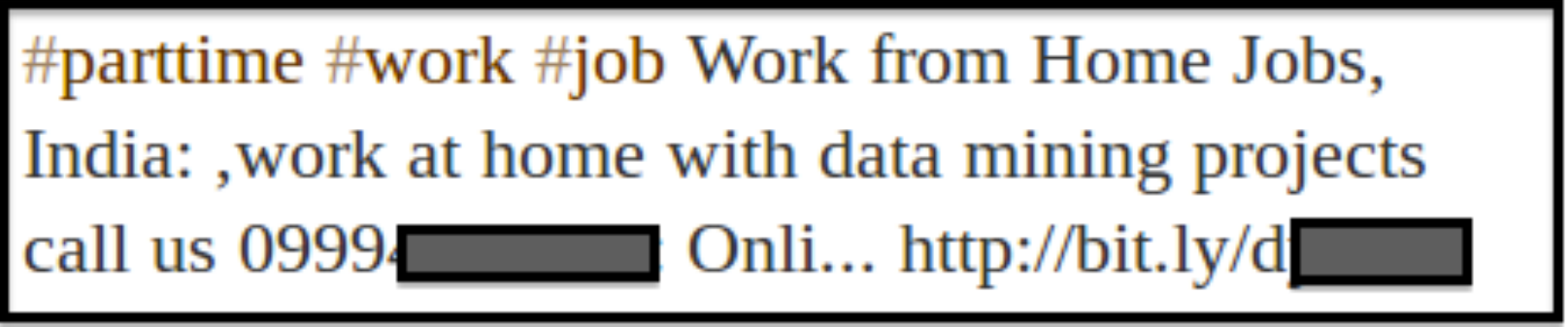}\label{fig:nodash}
   }
   \quad
   \subfigure[Pattern 2: One dash after country code]{
   \includegraphics[scale=0.3]{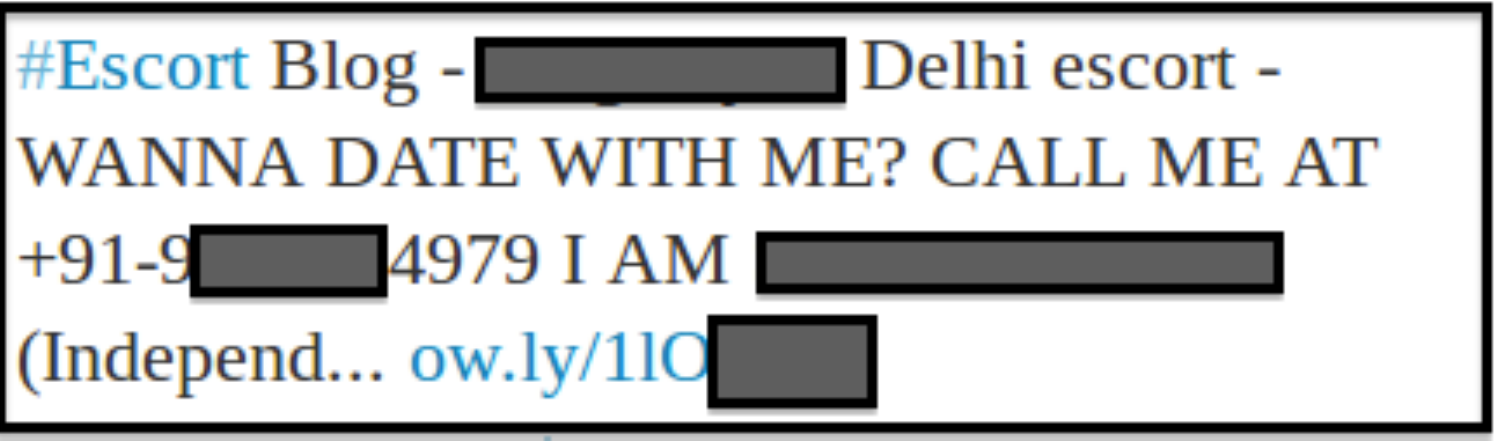}\label{fig:dash1}
   }
   \quad
      \subfigure[Pattern 3: Two dashes]{
   \includegraphics[scale=0.3]{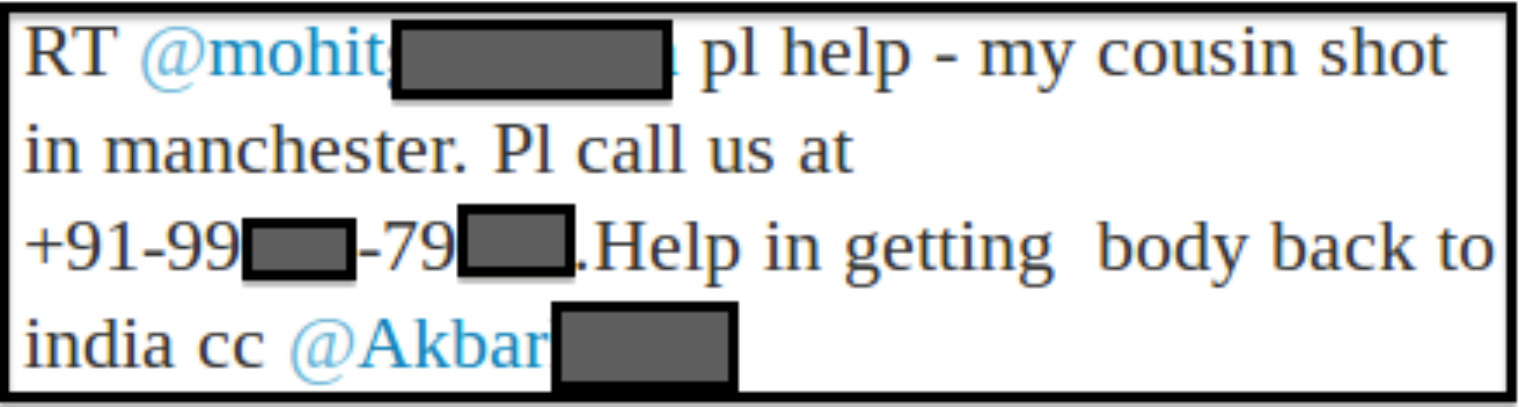}\label{fig:dash5}
}\quad
     \subfigure[Pattern 4: Three dashes]{
   \includegraphics[scale=0.3]{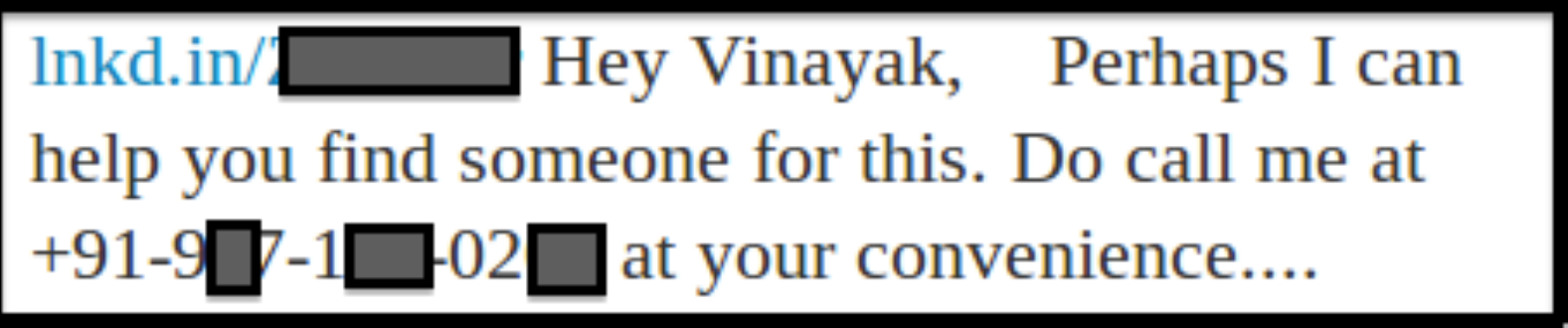}\label{fig:dash3}
   }
   \subfigure[Pattern 5: One space after country code]{
   \includegraphics[scale=0.3]{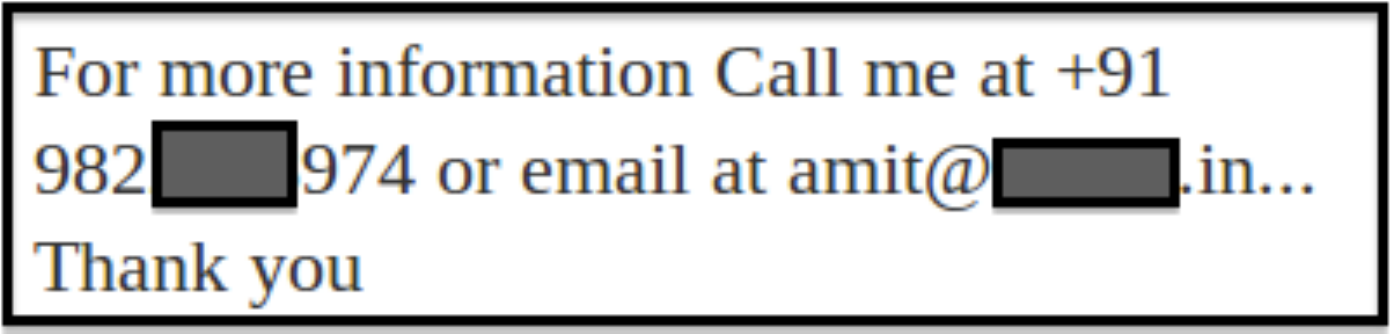}\label{fig:space1}
}
   \subfigure[Pattern 6: Two spaces]{
   \includegraphics[scale=0.3]{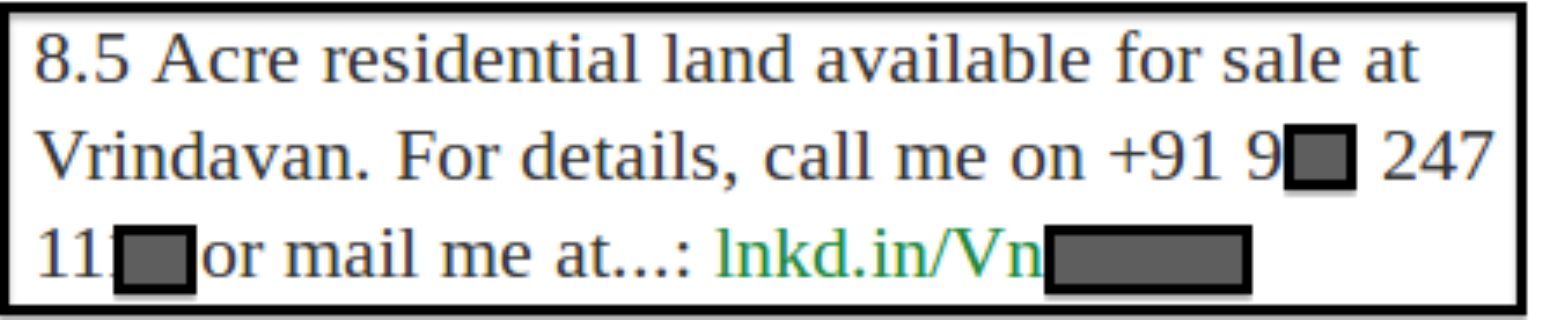}\label{fig:space3}
}
   \subfigure[Pattern 7: Dots between numbers]{
   \includegraphics[scale=0.3]{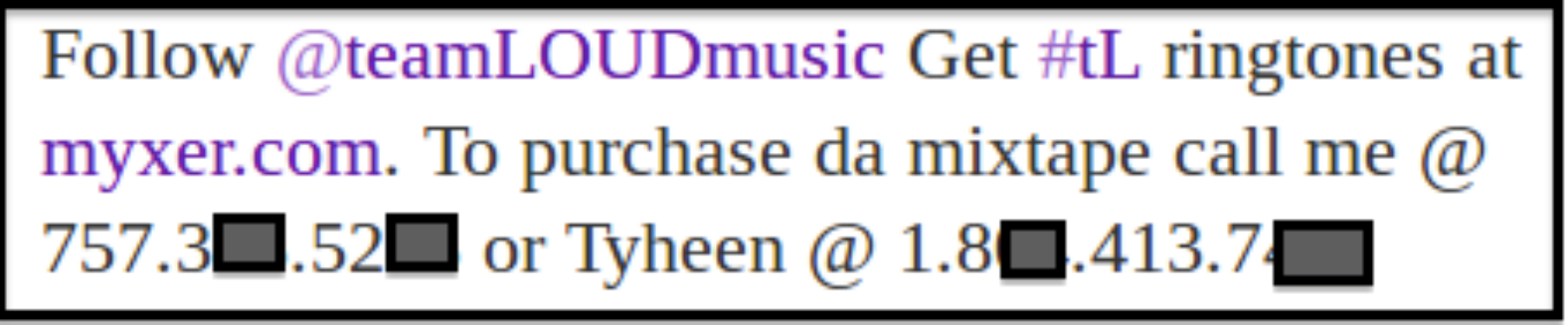}\label{fig:dots}
}

   \caption[Various formats and patterns in which users posted Indian mobile numbers on OSNs.]{Various formats and patterns in which users posted Indian mobile numbers on OSNs. Mobile numbers were prefixed with either trunk code `0' or country code `+91' while others had no prefix.}
      \label{fig:patterns}

\end{figure*}

\indent To tag Indian mobile numbers in users' posts and users' bio, we exploited the standard convention and structure of an Indian mobile number. It is a 10 digit number, where first digit should start with either 9 or 8 or 7. It can be prefixed with +91 or 0, where +91 is a country code and 0 is a trunk prefix.~\footnote{\url{http://www.dot.gov.in/sites/default/files/nnp2003.pdf}} We used rule-based named entity recognition~\cite{nadeau2007survey} and created a set of regular expression rules which captured Indian mobile number structure, to filter out Indian mobile numbers from posts and bio of users. We further observed that most users post Indian mobile numbers in different patterns (see Figure~\ref{fig:patterns}). We modified our regular expression rules to capture all possible ways of posting an Indian mobile number on social networks. We categorized Indian numbers prefixed with +91 as ``Category +91" numbers (see Figure~\ref{fig:dash1}), prefixed with 0 as ``Category 0" (see Figure~\ref{fig:nodash}), and prefixed with nothing as ``Category void" (see Figure~\ref{fig:dots}).
\pagebreak

Table~\ref{table:stats} shows the count of mobile numbers collected from tweets or bio on Twitter and public posts or names on Facebook.
\begin{table*}[!htb]
\begin{center}
\begin{tabular}{|c|l|l|l|l|l|l|}
\hline
 Numbers & \multicolumn{2}{c|}{\textbf{Category +91}}  & \multicolumn{2}{|c|}{\textbf{Category 0}} & \multicolumn{2}{|c|}{\textbf{Category void}} \\ \hline
\small{ [Till Apr 20, 2013]}& Twitter & Facebook  & Twitter & Facebook & Twitter & Facebook \\ \hline
\textbf{Mobile numbers} & 885 & 2,191 & 14,909 & 8,873 & 25,566 & 25,294 \\ \hline
\textbf{User profiles} & 1,074 & 2,663 & 17,913 & 9,028 & 31,149 & 25,406 \\ \hline
\end{tabular}
\end{center}
\end{table*}

\begin{table*}[!htb]
\vspace{-10mm}
\begin{center}
\begin{tabular}{|c|l|l|}
\hline
Numbers &\multicolumn{2}{|c|}{\textbf{Total}} \\ \hline
&Twitter & Facebook\\ \hline
\textbf{Mobile numbers} & 41,360 & 36,358  \\ \hline
\textbf{User profiles} & 49,817 & 36,588 \\ \hline
\end{tabular}
\end{center}
\vspace{-2mm}
\caption{Descriptive statistics of the mobile numbers collected from Twitter and Facebook.}
\vspace{-2mm}
\label{table:stats}
\end{table*}

\section{Data validation}
\indent Rule-based named entity recognition used to extract Indian mobile numbers from public posts and bio in the earlier stage, relied on a set of regular expressions. Regular expressions incorporated Indian mobile number structure and therefore misinterpreted certain other country number strings as Indian mobile numbers.

\indent Mobile number format for few countries (United Kingdom,~\footnote{\url{http://stakeholders.ofcom.org.uk/binaries/telecoms/numbering/Numbering_Plan_October_2013.pdf}} Indonesia~\footnote{\url{http://en.wikipedia.org/wiki/Telephone_numbers_in_Indonesia}} and USA~\footnote{\url{http://www.nanpa.com/enas/npaDialingPlansReport.do}}) is similar to that of an Indian mobile number. UK mobile numbers are also 10-digit numbers starting with 07, which were confused as Indian mobile numbers prefixed with 0 and starting with 7. Few Indonesian mobile numbers are prefixed with 0881 / 0882 followed by a 7-digit mobile number (depending on the operator e.g. PT Sinar Mas) making it a 10-digit mobile number, similar to an Indian mobile number prefixed with 0 and starting with 8. USA mobile numbers also follow 10-digit format with first three digits representing area code, ranging from 2-9, therefore USA mobile numbers without country code and with area codes starting with 7, 8, 9 are similar to an Indian mobile number. Moreover, Category 0 and Category void numbers do not follow the international mobile numbering format. Hence, we made them pass through a Data Validation stage.

\indent In Data Validation stage we used a service from ~\url{http://trackmobileonline.co.in}, which checked if a number's first four digits belonged to a valid Indian mobile number series. We observed that 19,934 mobile numbers out of 23,405 in Category 0 (85\%), and 42,360 numbers out of 49,946 in Category void (85\%), were marked as  Indian mobile numbers by the service (see Table~\ref{table:Validation}).

\begin{table}[h!]
\begin{center}
\begin{tabular}{|p{2cm}|l|l|l|l|}
\hline

\textbf{Numbers}& \multicolumn{2}{|c|}{ \textbf{Category 0}} & \multicolumn{2}{|c|}{\textbf{Category void}} \\ \hline
& Passed & Failed  & Passed & Failed  \\ \hline
\textbf{Twitter} & 12,681 & 2,228 & 21,443 & 4,123  \\ \hline
\textbf{Facebook} & 7,586 & 1,287 & 21,715 & 3,579 \\ \hline
\textbf{Total} & 20,267 & 3,515 & 43,158 & 7,702  \\ \hline

\end{tabular}
\end{center}
\caption{Statistics of valid Category 0 and Category void Indian mobile numbers}
\label{table:Validation}
\vspace{-3mm}
\end{table}

\indent On manual verification, we observed some non-Indian numbers were marked as `Passed' by the service. 
In order to avoid any bias or noisy inference by including Category 0 and Category void numbers we considered \emph{only} Category +91 mobile numbers for our analysis, since they were confirmed to be Indian mobile numbers (follow International format). We keep validation and analysis of Category 0 and Category void numbers as future work of this study.

\chapter{Analysis}\label{chapter:analysis}
\onehalfspacing
In this section, we present detailed analysis to understand mobile number sharing behavior on OSNs.
\section{Context analysis} \label{Context}
We first attempt to comprehend the contexts in which mobile numbers were shared on OSNs. To understand the context, we extracted most frequent words from the bio descriptions and collected posts which shared the number. We removed stop words and performed stemming~\cite{PorterStemmer}
to avoid repeated forms of the same root word. Words were then supplied to a text analysis software, LIWC.~\footnote{\url{http://www.liwc.net/}} LIWC returned the category associated with each word as well as ranking of the categories to which most words belonged to.

Top-5 LIWC categories ranked in order of popularity on Twitter were `social', `affect', `cogmech', `work', `leisure' while on Facebook `work', `cogmech', `social', `time' and `space' were the top ones. `Work' category (words such as \emph{office, computer}) was more popular on Facebook than Twitter, while `leisure', and `social' category (words such as \emph{travel, apartments, music}) were more popular on Twitter than Facebook. `Time' category was popular on both OSNs, however in different contexts. On Twitter, `time' category marked words such as \emph{urgency, need, now} while on Facebook, `time' category marked official context words \emph{such as year, time, weekday.}
\begin{figure*}[t] 
   \centering
   \subfigure[Twitter Tag-cloud]{
   \includegraphics[scale=0.25]{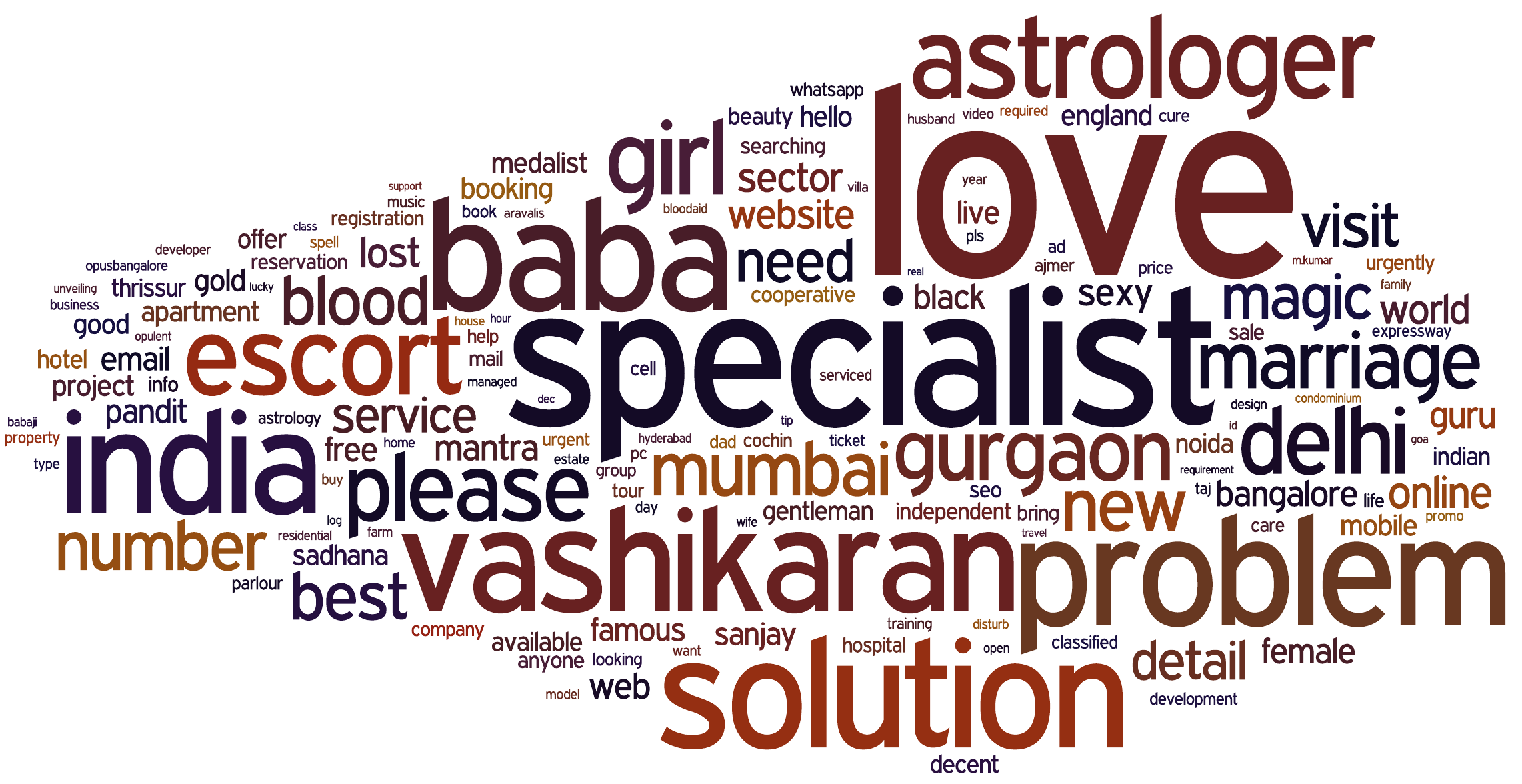}\label{fig:tweet_tag}
   }
\quad
   \subfigure[Facebook Tag-cloud]{
   \includegraphics[scale=0.25]{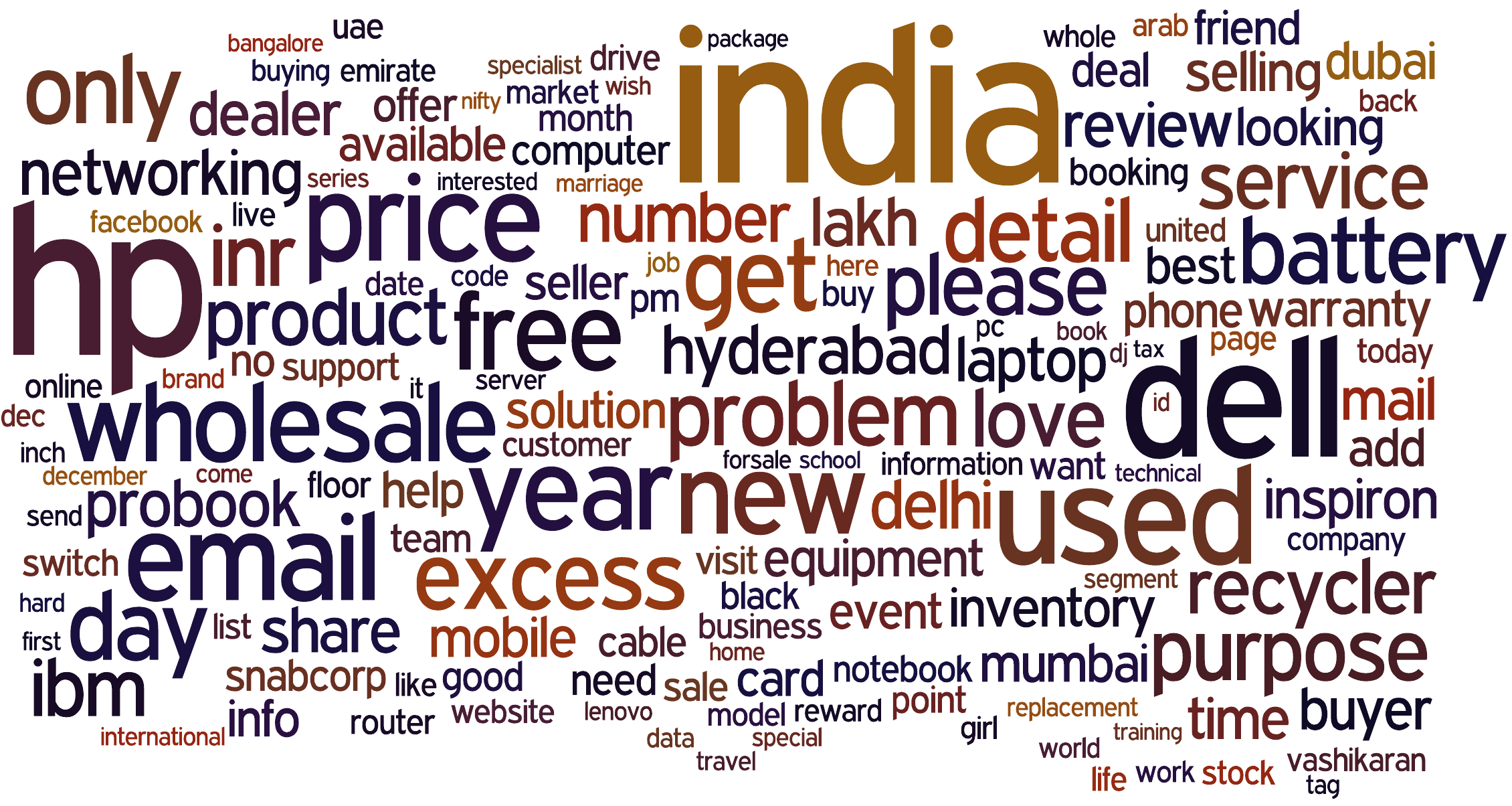}\label{fig:fb_tag}
   }
   \quad
     \subfigure[Twitter Bio Tag-cloud]{
   \includegraphics[scale=0.25]{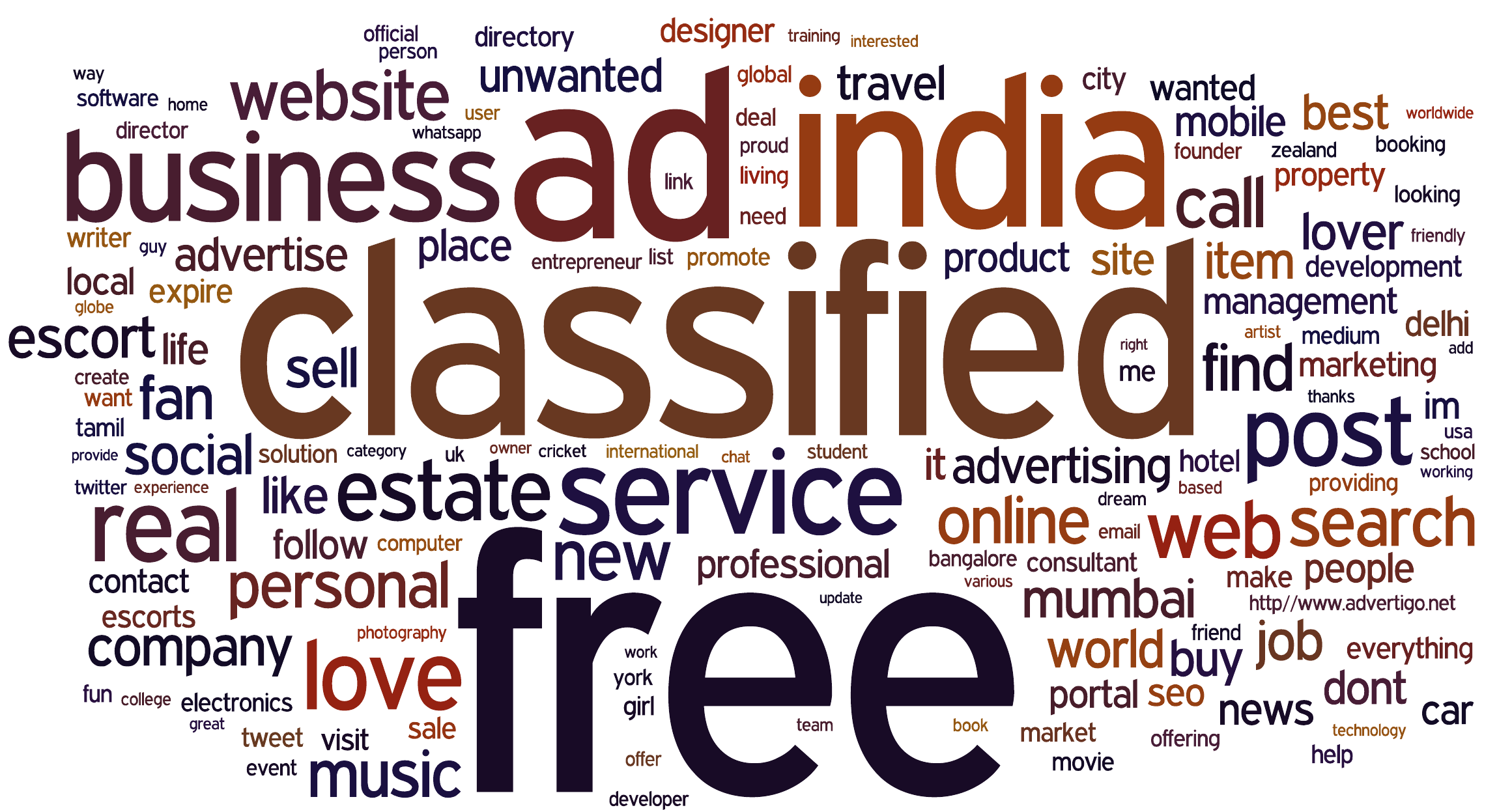}\label{fig:bio_tweet_tag}

   }
   \quad
   \subfigure[Facebook Name Tag-cloud]{
   \includegraphics[scale=0.33]{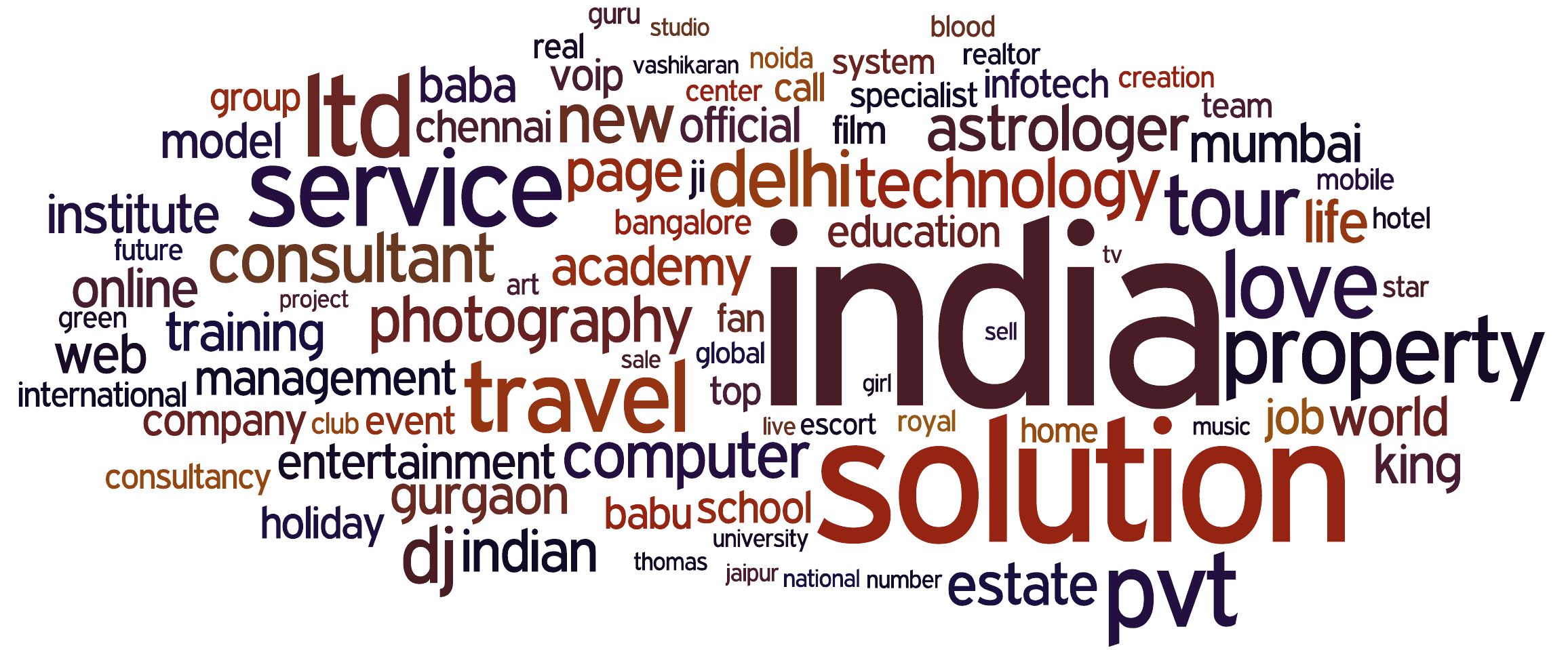}\label{fig:username_fb_tag}
}
   \caption[Figure \ref{fig:tweet_tag} and \ref{fig:fb_tag} shows the context in which users leaked mobile numbers on Twitter and Facebook.]{Figure \ref{fig:tweet_tag} and \ref{fig:fb_tag} shows the context in which users leaked mobile numbers on Twitter and Facebook.
   Users posted mobile numbers in context of emergency, escort services, entertainment, marketing on Twitter while in context of IT facilities on Facebook. Figure~\ref{fig:bio_tweet_tag} and \ref{fig:username_fb_tag} shows the genre of users who shared Indian mobile number on Twitter and Facebook. }
   \label{fig:why}
\end{figure*}
\indent To clearly understand the contexts, we manually analyzed word-clouds of the most frequent words (see Figure~\ref{fig:tweet_tag} and \ref{fig:fb_tag}). We observed words such as \emph{blood, specialist, hospital, love, sexy, escort, girl, music, movie, fun, offer, reservation, ticket, hotel, seo, sale, astrologer, business} in Figure~\ref{fig:tweet_tag}. We infer that on Twitter, users post Indian mobile numbers, majorly to ask for blood donations / aid, help in emergency situations, to promote escort business, to promote entertainment, to market for travel, holiday, hotel packages, and to buy / sell products, etc. Such a behavior is understandable since Twitter acts as a news media, and marketing platform for most companies ~\cite{Kwak10www}. On Facebook, users posted Indian mobile numbers majorly in context of Information Technology (IT) facilities and education related products, evident by the presence of words such as \emph{price, hp, battery, dell, laptop, ibm, email, notebook, computer }(see Figure~\ref{fig:fb_tag}). We infer that users post mobile numbers on social media platforms in order to benefit from social network structure and promote their business by spreading the contact information (mobile number) to large number of users.

\section{Genre analysis}
Context analysis gave an understanding that mobile numbers were leaked in different contexts, however profession of users who predominantly posted mobile numbers is unclear. We now attempt to understand the profession of the users who posted mobile numbers.
Since, there is no explicit attribute mentioning profession of a user on Twitter,~\footnote{\url{https://dev.twitter.com/docs/api/1}} we created a word-cloud of the description (bio) of the users,
who shared mobile numbers either via public posts or in their bio (see Figure~\ref{fig:bio_tweet_tag}). We observed words as \emph{classified, free, ad, service, business, estate, job, escort, company} and infer that Twitter is heavily used by advertising sites, marketers and business-oriented users and they intend to expand their business by enhancing reachability via mobile numbers. Profession attribute of a user is not available publicly via Facebook API ~\footnote{\url{https://developers.facebook.com/docs/reference/api/user/}} as well, so we created a word-cloud of names of users who have shared at least one Indian mobile number via their public post (see Figure~\ref{fig:username_fb_tag}). We chose names because no other attribute on Facebook was publicly available with which we could extract profession of the users. We observe words as \emph{solution, computer, technology, education, institute, website, training, management, admission, helpline, love, astrologer, travel, photography, entertainment} and infer that technology experts, academicians, marketers and artists majorly post mobile numbers on Facebook.

\section{Gender distribution}
Privacy studies show that higher proportion of females (19\%) consider mobile number as PII as compared to males (10\%) in India~\cite{kumaraguru:privacy-in-india:-attitud:2012:yuqfj}.
To understand if females were privacy conscious while sharing mobile number on OSNs, we observed gender of the users who posted Indian mobile numbers. We extracted gender of all the Facebook users using Facebook Graph API. Since Twitter API does not have gender as profile attribute, we derived gender for only those Twitter users who mentioned their Facebook identity on Twitter's URL attribute.
Table~\ref{who:gender} shows detailed statistics of gender of users who shared mobile numbers. Less than 20\% female users leaked mobile numbers publicly on OSNs. This implies that females are conservative while sharing mobile numbers on OSNs, inline to the observation in~\cite{kumaraguru:privacy-in-india:-attitud:2012:yuqfj}.

\begin{table}[ht]

\begin{center}
\begin{tabular}{|p{3.6cm}|p{1.8cm}|p{1.8cm}|}
\hline
 & \raggedright{{\bf Facebook}} & {\bf{Twitter}} \\ \hline
Total users & 2,663 & 1,074  \\
\hline
\raggedright{Gender available (G)} & 1,438 & 29 \\ \hline
Females (F) & 220  & 6 \\ \hline
Males (M) & 1,218  & 23\\ \hline
Percentage (F / G) &  \textbf{15\%} & \textbf{20\%} \\ \hline
\end{tabular}
\caption{Gender distribution. Few females leak mobile numbers on OSNs as compared to males. }
\label{who:gender}
\end{center}
\end{table}

\vspace{-6mm}
\section{Ownership analysis}\label{section:owner_analysis}
\begin{figure*}[t] 
   \centering

   \includegraphics[scale=0.7]{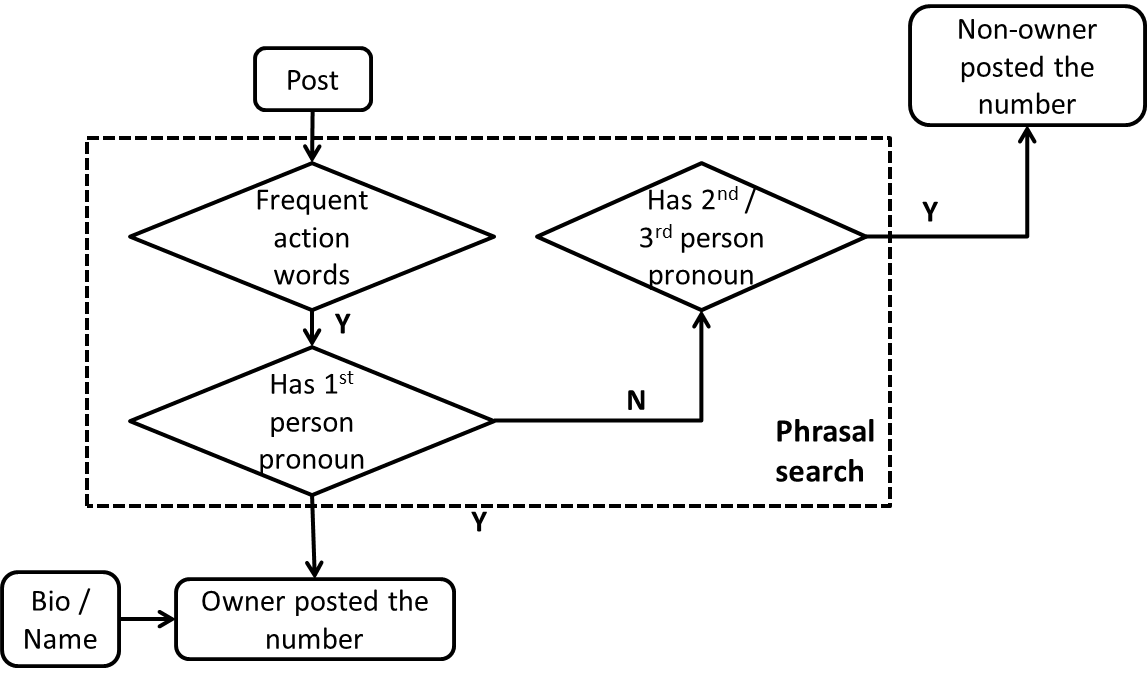}

   \caption{Ownership Analysis Technique}
   \label{fig:own_tech}
\end{figure*}

Exposure of mobile numbers by non-owners might lead to unwanted privacy leaks and annoyance to their owners.~\footnote{\url{http://thenextweb.com/media/2011/07/10/supposed-phone-number-of-news-internationals-chief-executive-leaked-on-twitter}} We therefore analyzed weather the mobile number owner himself posted his number at the first place or someone else posted his number on online social network. For this we used methodology shown in Figure ~\ref{fig:own_tech}. 

\begin{figure*}[t] 
   \centering
   \subfigure[Mobile number posted by the owner of mobile number.]{
   \includegraphics[scale=0.25]{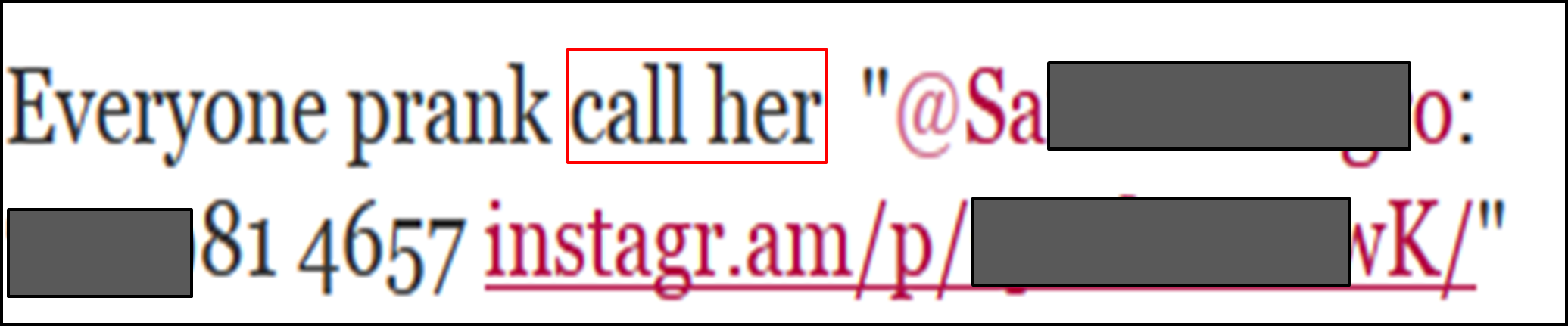}\label{fig:post_non}
   }
\quad
   \subfigure[Mobile number posted by the non-owner of mobile number. ]{
   \includegraphics[scale=0.25]{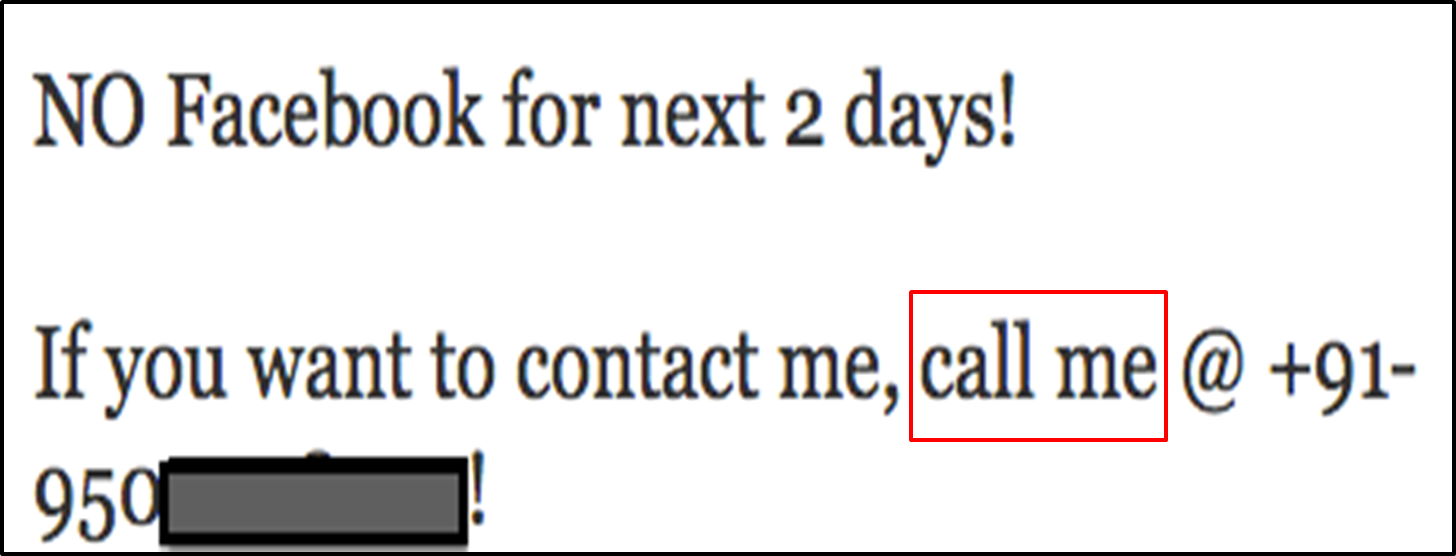}\label{fig:post_own}
   }

   \caption{Sample posts where mobile number is shared by ~\ref{fig:post_non} non-owner of the shared mobile number and ~\ref{fig:post_own} owner of the shared mobile number }

\end{figure*}

For each mobile number collected from Twitter (885) and Facebook (2,191), we retrieved the first tweet (post) from our dataset, sharing that mobile number on Twitter (or Facebook). The mobile number was marked as  `leaked by its owner', if the tweet (post) included a first person pronoun such as \emph{me, my, us, mera (my in English)} along with most frequent action verbs such as \emph{call, text, sms, ping, whatsapp, message, contact}. For instance we checked for the presence of phrases like - ``call us", ``text us", ``my contact". Figure ~\ref{fig:post_own} shows an example post where owner himself leaked his mobile number, according to our detection technique. The mobile number was marked as `leaked by a non-owner', if the tweet (post) included second person pronoun such as \emph{you, your, yours} or third person pronoun such as \emph{his, her, them} along with same action verbs as used with first person pronouns before. Figure ~\ref{fig:post_non} shows an example post of user (non-owner) leaking some other person's mobile number, according to our detection technique. Researchers used only pronouns to check for ownership~\cite{Mao:2011:LTA:2046556.2046558}, this may give false positives like - ``You may call me at xxx", however we avoided it by using phrases here. We compared the two algorithms, the results are presented toward the end of this section. We also assumed that mobile numbers shared on Twitter via bio or on Facebook via name are users' own mobile numbers.

Table~\ref{tab:owner} shows the descriptive statistics of mobile numbers which were leaked by their owners and non-owners.  Two hundred and ninety one mobile numbers (32.8\%) were shared by their owners while only 18 mobile numbers (2.0\%) were shared by non-owners on Twitter. Four hundred and eighty seven mobile numbers (22\%) were shared by owners, and 25 mobile numbers (1.1\%) were shared by non-owners on Facebook. Example post where owner shared his mobile number is {\emph {``F1 INR 2500/- tickets are available with me..!! Limited stocks..!! Ping me or call me up on +91 989 xxx xxxx asap!}"} Example post where non-owner shared the mobile number is \emph{``@VodafoneIN My friend Debasrita took a new connection (+91-73816xxxxx), she is having issues. Please contact her at +91-9556xxxxxx"}. Exposure of mobile numbers by non-owners might lead to unwanted privacy leaks and annoyance to their owners~\cite{Non-ownerLeak}. For remaining mobile numbers, the methodology used could not infer if the numbers were shared by the owners or non-owners. Example post is ``\emph{Need a male punjabi artist of age 35 for a ad in \#chennai pls contact +91 98-41-xxxxxx}."

\begin{table}[!h]
\begin{center}
\begin{tabular}{|p{3.7cm}|p{3.5cm}|p{2cm}|} \hline
\textbf{Social Network - Source of leakage} &\textbf{ Mechanism} & \textbf{\# Mobile numbers} \\ \hline
\multirow{2}{*}{Twitter - Owner} & Bio & 155  \\
& Tweet & 136 \\ \hline
\multirow{1}{*}{Twitter - Non-owner} & Tweet & 18 \\  \hline
\multirow{2}{*}{Facebook - Owner} & Message & 422  \\
& Caption & 38\\
& Description & 8 \\
& Name & 17 \\
& Others & 2 \\ \hline
\multirow{1}{*}{Facebook - Non-owner} & Message &  25\\

 \hline
\end{tabular}
\end{center}
\caption[Statistics of mobile numbers shared by owners and non-owners on Twitter and Facebook.]{Mobile numbers shared by owners and non-owners on Twitter and Facebook. Most mobile numbers were leaked by owners themselves, though few were leaked by non-owners. }
\label{tab:owner}
\end{table}

\indent In the past, researchers used only pronouns to check for ownership
~\cite{Mao:2011:LTA:2046556.2046558} however in our algorithm we used phrasal level search
~\ref{section:owner_analysis}. We now compare the performance of the two methods in detecting if the mobile number owner shared his number himself or someone else shared it. We evaluate both methods by comparing their results with human annotated ground truth dataset. The ground truth dataset was prepared by manually tagging all the posts sharing mobile numbers in the dataset, which also has a pronoun (basic condition needed by both methods to work). The posts were tagged in accordance with - if they have mobile number(s) `shared by owner', or `shared by non-owner'. For some posts, it was hard for a human to understand if they were posted by the owner or non-owner, such posts were tagged as - `ownership not clear'. Example posts falling under the category - `ownership not clear' were ``@VodafoneIN +91843xxxxxxx time of call 14:51 today and purpose ICICI bank credit card offer ... Can we check this ???", ``@rusXXXXa oops sorry. I had just woken up. This is the correct number - +9198xxxxxxx4".

\indent We used false positive rate as a measure to compare the two methodologies. False positive rate is the fraction of mobile numbers marked incorrectly by a method to the total mobile numbers tagged by the same method. We report the result only for numbers whose ownership could be disambiguated on manual tagging (they are not tagged as `ownership not clear'), on Twitter we had 200 such numbers however on Facebook there were 861 such numbers. On Twitter, we observed that the previous methodologies~\cite{Mao:2011:LTA:2046556.2046558} had a false positive rate of 0.08\% (16/200) and our methodology~\ref{section:owner_analysis} had a false positive rate of 0.01\% (2/154). On Facebook, we observed that the previous methodologies~\cite{Mao:2011:LTA:2046556.2046558} had a false positive rate of 0.13\% (109/861) and our methodology~\ref{section:owner_analysis} had a false positive rate of 0.04\% (19/495). The methodology we proposed~\ref{section:owner_analysis} clearly performed better on both the OSNs, by having a lower false positive rate.


\section{Topographical distribution}

We probe into the location of the users, who shared Indian mobile numbers on OSNs, to understand if users of few locations more actively posted mobile numbers on social networks than others.  We analyzed geo-tagged posts which shared mobile numbers on both social networks. We identified only 13 geo-tagged tweets on Twitter, where 9 unique users shared 12 mobile numbers, listed in Table~\ref{location_bio}. We did not find any geo-tags in Facebook public posts which shared the number.

\begin{table*}[ht]
\raggedright{
\begin{tabular}{|l|p{4cm}|p{4cm}|p{4.5cm}|}

\hline
 & \raggedright{{\bf {Location of user}}} & {\bf{Location via bio of user}} &  {\bf{Location via geo-tagged posts}} \\ \hline
\textbf{Country} & India, United States, United Kingdom, United Arabs, Canada & United States, India, Russia, Belgium, Canada & India  \\
\hline
\textbf{State} & Maharashtra, Karnataka, Delhi, Tamil Nadu, Andhra Pradesh  & Maharashtra, Texas, Pennsylvania, Moscow Federal City, California &  Maharashtra, Karnataka, Tamil Nadu, Delhi \\ \hline
\textbf{City} &  Delhi, Mumbai, Bangalore, Chennai, Hyderabad  & \raggedright{Moscow, Mumbai, Fallowfield, Taipei City, Addison} & Delhi, Mumbai, Mysore, Coimbatore \\ \hline
 \end{tabular}
\caption{Top states and cities, from where most Indian mobile numbers were shared on OSNs.}
\label{location_bio}
}\end{table*}


With few geo-tagged tweets, we investigated whether location of the users who shared mobile numbers can be estimated either via their `location' attribute or bio description on Twitter~\cite{pontes:beware-of-what-you-share::2012:yuqfj}. We used Yahoo Maps~\footnote{\url{http://developer.yahoo.com/maps/}} to trace a location, present in users' location attribute or bio attribute, to a city, state and country. We found location of 780 users via `location' attribute and of 753 users via their bio description. We ignored locations which did not map to real geographical locations like ``Justin Beiber's heart". Table~\ref{location_bio} shows the country, state and city from where most Indian mobile numbers were shared, either via `location' attribute or bio description. We infer that mobile numbers were largely shared by users of urban cities in India. 
%
%

\indent We further looked at the location from where the leaked mobile numbers were issued. Indian mobile numbers can be splitted up into 2 parts – first part (1st four digits) represents the network operator of which the number is part of and second part (last six digits) represents the subscriber number. The first part can be used to infer the telecom zone/circle and hence the location from where the number was issued. These telecom zones/circles are categorized into Metropolitan / A / B / C circle. Metro circles are the ones with High population density. `A' circle has largest subscriber base, then `B' circle and least is in `C' circle.~\footnote{host.comsoc.org/sistersocieties/india\_iete/circles.pdf} With this background we looked at the issue location of leaked mobile numbers (see Table~\ref{table:Number_circle}). We found that most numbers belonged to Indian metropolitan telecom circles and large cities `A' circles. We also observed that smaller circles like Punjab and Andhra Pradesh post more mobile numbers in comparison to most of the bigger circles.


\begin{table}[!ht]
\begin{center}
\begin{tabular}{|p{4cm}|p{2.5cm}|p{2.5cm}|}
\hline
{\bf Telecom circle}	&\textbf{Category}& {\bf \# of mobile numbers}\\
\hline
Delhi & Metropolitan	&582\\
\hline
Mumbai & Metropolitan &312\\
\hline
Karnataka	 &``A" Circle & 233\\
\hline
Punjab &``B" Circle 	&226\\
\hline
Rajasthan &``B" Circle  & 171 \\
\hline
Andhra Pradesh &``A" Circle 	&164\\
\hline
Kerala &``B" Circle 	&158\\
\hline
Maharashtra  &``A" Circle  & 140 \\
\hline
Gujrat &``A" Circle 	&135\\
\hline
Tamil Nadu & ``A" Circle 	&102\\
\hline
Kolkata 	&``B" Circle 	&  75\\
\hline
Uttar Pradesh (West)	&``B" Circle 	&75\\
\end{tabular}

\begin{tabular}{|p{4cm}|p{2.5cm}|p{2.5cm}|}
\hline
Uttar Pradesh (East)	&``B" Circle 	&70\\
\hline
Madhya Pradesh	&``B" Circle 	& 64\\
\hline
Haryana	&``B" Circle 	&37\\
\hline
Bihar $\&$ Jharkhand	&``C" Circle 	&35\\
\hline
West Bengal	&``B" Circle 	&31\\
\hline
Assam	&``C" Circle 	&28\\
\hline
Jammu $\&$ Kashmir	&``C" Circle 	&27\\
\hline
Himachal Pradesh	&``C" Circle 	&18\\
\hline
Orrisa	&``C" Circle 	&16\\
\hline
North East	&``C" Circle 	&8\\
\hline
\end{tabular}
\caption[Telecom circle and count of mobile numbers associated with the circle in India.]{Telecom circle and count of mobile numbers associated with the circle in India. Most mobile numbers which were exposed on OSNs, belonged to populated telecom circles. }
\label{table:Number_circle}
\end{center}
\end{table}

%

We hypothesize that users of cities with higher Internet user base~\cite{IAMAI} are most active in sharing mobile numbers on OSNs. To test the hypothesis, we calculated correlation between top-8 Indian cities ranked by active Internet users~\cite{IAMAI} and top-8 Indian cities ranked by number of mobile numbers shared on Twitter (location extracted from `location' attribute as well as bio of a user). We observed a positive correlation of 0.86, therefore we conclude that penetration of Internet at an Indian location impacts number of mobile numbers shared from that location on social media.

We present a lists of top 5 Telecom Operator and corresponding number of Category +91 mobile numbers taking service from the operator in Table \ref{table:Number_operator}. The ranking corresponds to the top 5 operators (in accordance to subscriber base), ~\footnote{\url{http://www.coai.com/Uploads/MediaTypes/Documents/All-India-GSM-figures-April-2013.xls}} however Reliance has a larger subscriber base than Vodafone.
\begin{table}[!ht]
\begin{center}
\begin{tabular}{|p{6cm}|p{2cm}|}
\hline
{\bf Operator}	&{\bf Frequency}\\
\hline
Airtel 	&845\\
\hline
Vodafone	&772\\
\hline
Reliance communications	&255\\
\hline
Idea & 223\\
\hline
BSNL	&182\\
\hline
\end{tabular}
\caption{Top 5 Mobile Number Operator of Category +91 numbers}
\label{table:Number_operator}
\end{center}
\end{table}

\section{Source analysis}
\begin{figure}
  \centering
  \includegraphics[scale=0.5]{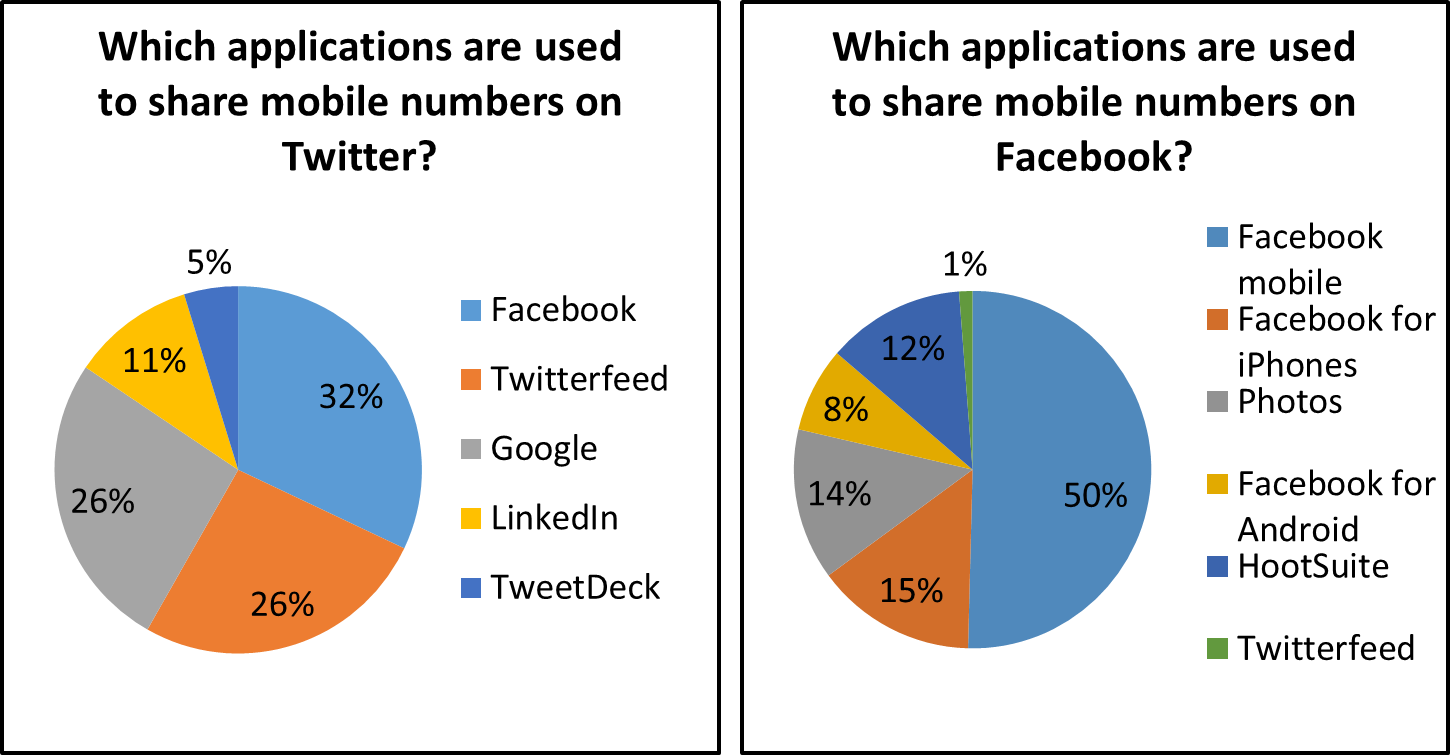}
  \caption[Figure shows the devices / tools / apps used to post mobile numbers on social networking sites.]{Figure shows the devices / tools / apps used to post mobile numbers on social networking sites. We observed other OSNs like Facebook, Google and LinkedIn are used to share mobile numbers, implying users are not restricted to one social network, but are posting mobile numbers on multiple networks. We also observed usage of 3rd party applications like Twitterfeed, which might be used to post on multiple OSNs together, hence strengthening our argument.}\label{fig:source}
\end{figure}

Same mobile numbers posted from profiles coming from different social networks can be used to find related accounts across social networks. To understand weather same number is posted across different Social Networks, we inquired the source or application by which most mobile numbers were posted on Twitter and Facebook (see Figure ~\ref{fig:source}).
To extract application used to post the number, we extracted `source' attribute of the tweet, available from Twitter API,~\footnote{\url{dev.twitter.com/docs/platform-objects/tweets}} and `application' attribute of the post, available from Facebook Graph API.~\footnote{\url{developers.facebook.com/docs/reference/api/post/}} On Twitter, apart from the web (234), mobile numbers were largely posted from social aggregators and other social networks such as Facebook (148), Twitterfeed (121), Google (121), LinkedIn (50), TweetDeck (22).  We observed major use of social aggregators and other social networks to post mobile numbers on Twitter indicating users might be sharing same mobile number not only on one OSN but on multiple OSNs simultaneously. On Facebook, most numbers were posted by Facebook mobile applications such as Facebook mobile (125), Facebook for iPhone (36 numbers), Photos (34),  Facebook for Android (19), and few by social aggregators such as HootSuite (31), and Twitterfeed (3). We observed major use of OS based Facebook mobile applications to post numbers on Facebook with comparatively less exploitation of social aggregators.

We infer that users push the same information across multiple OSNs in order to increase audience receiving the information. 
Out of 2,996 unique Category +91 numbers in our dataset, we found 80 mobile numbers were shared on both Twitter and Facebook.
Note that, same mobile number can be posted by many users using different applications.

\section{Network analysis}
To investigate if popularizing actions like posting the same number on multiple OSNs were effective, we analyzed networks formed by the dissemination of numbers on 
Twitter. Since Facebook Graph API does not provide public access to a user's friends / networks, we analyzed networks for mobile numbers shared on Twitter.
We constructed two networks, \textit{audience network} and \textit{dissemination network}. We define \textit{audience network} of a mobile number as a directed graph $ G_A (V_A, E_A)$, where nodes represent two kinds of users -- publishers and consumers. \emph{Publishers} are users who shared the mobile number by posting or retweeting it, colored as black. \emph{Consumers} are users who received the mobile number in their timeline by following the publishers, colored as orange. Edges represent the follower relationships between publisher and consumer, colored as orange, labelled as 2. Follower relationship implies that a consumer is a follower of a publisher (edge directed towards the follower). Using $ G_A$, one can analyze the nature and volume of the audience users to whom the mobile number was presented in their timeline. \textit{Dissemination network} of a mobile number is a directed graph $ G_D (V_D, E_D) $ where nodes represent only publishers of the mobile number and edges represent two kinds of relationships -- follower and retweet. Follower relationship implies that one publisher is a follower of other publisher, colored as black, labelled as 1. Retweet relationship implies that one publisher retweets the post of other publisher (edge directed towards the retweet publisher), colored as magenta, labelled as 3. $G_D$ helps in understanding the diffusion patterns of a mobile number among its publishers.

\begin{figure*}[!t] 
    \centering
   \subfigure[Emergency: Audience network]{
   \includegraphics[scale=0.2]{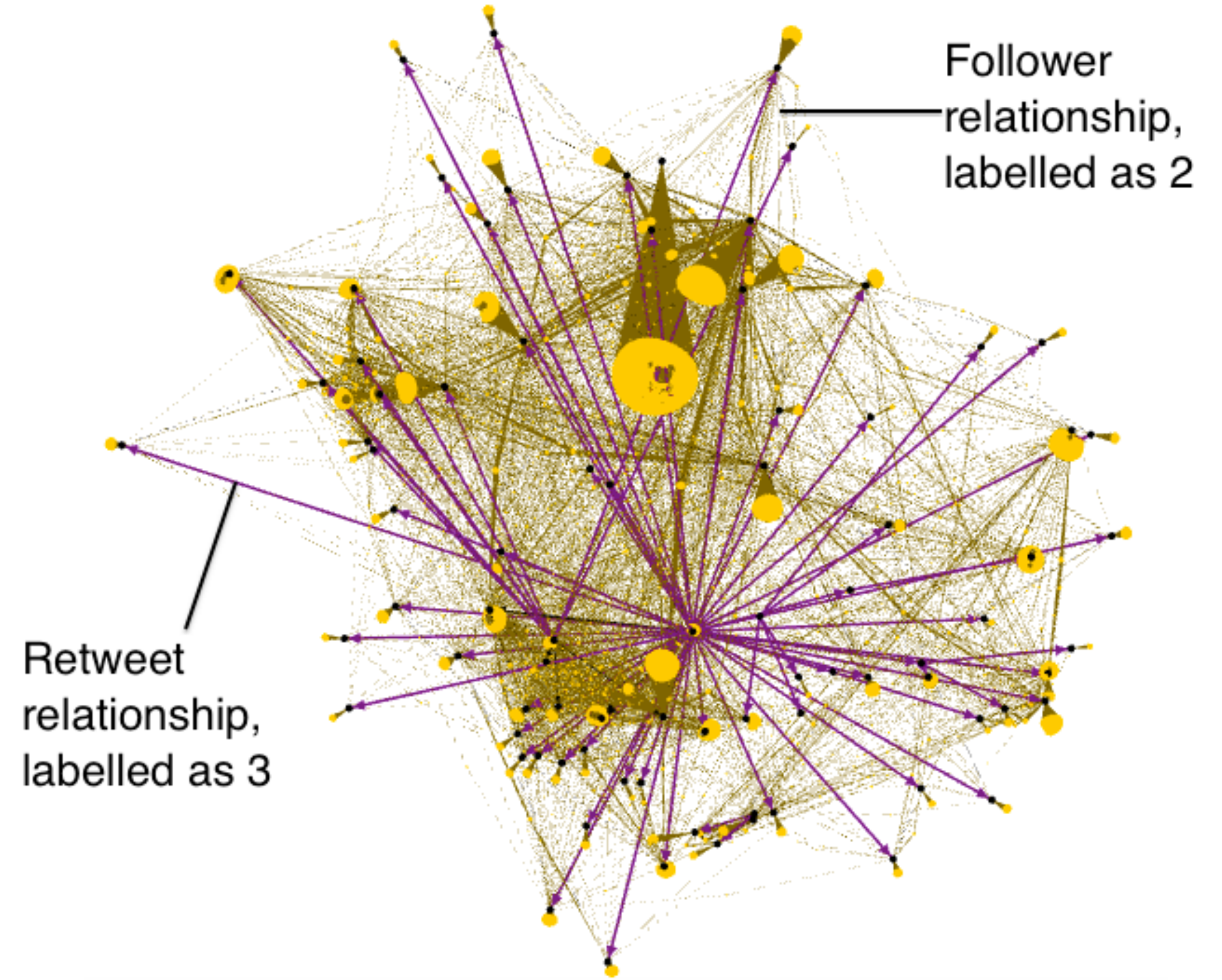}\label{fig:emergency_a}
   }
   \quad
      \subfigure[Entertainment: Audience network]{
   \includegraphics[scale=0.2]{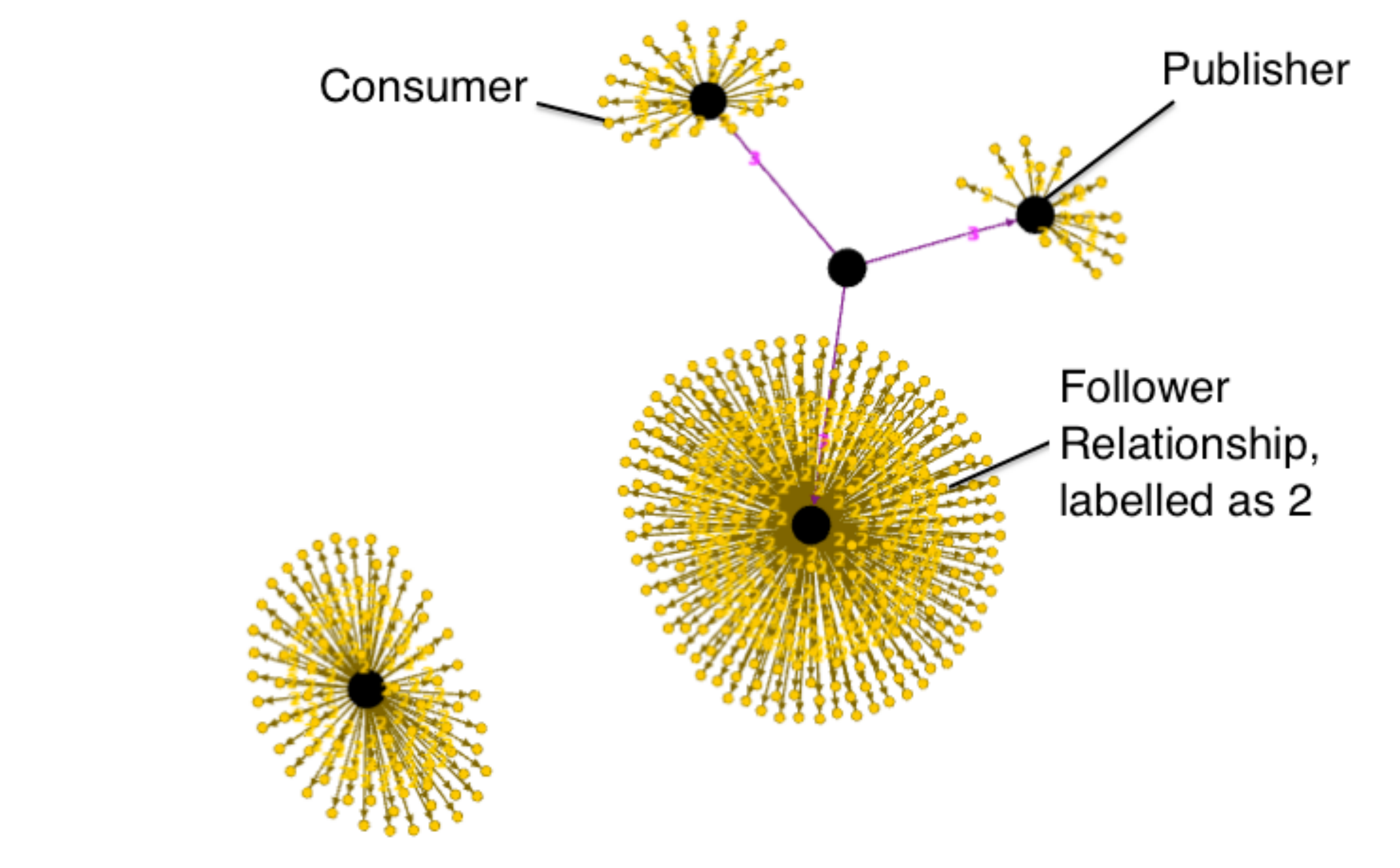}\label{fig:fun_a}
}
\quad
    \subfigure[Marketing: Audience network]{
   \includegraphics[scale=0.2]{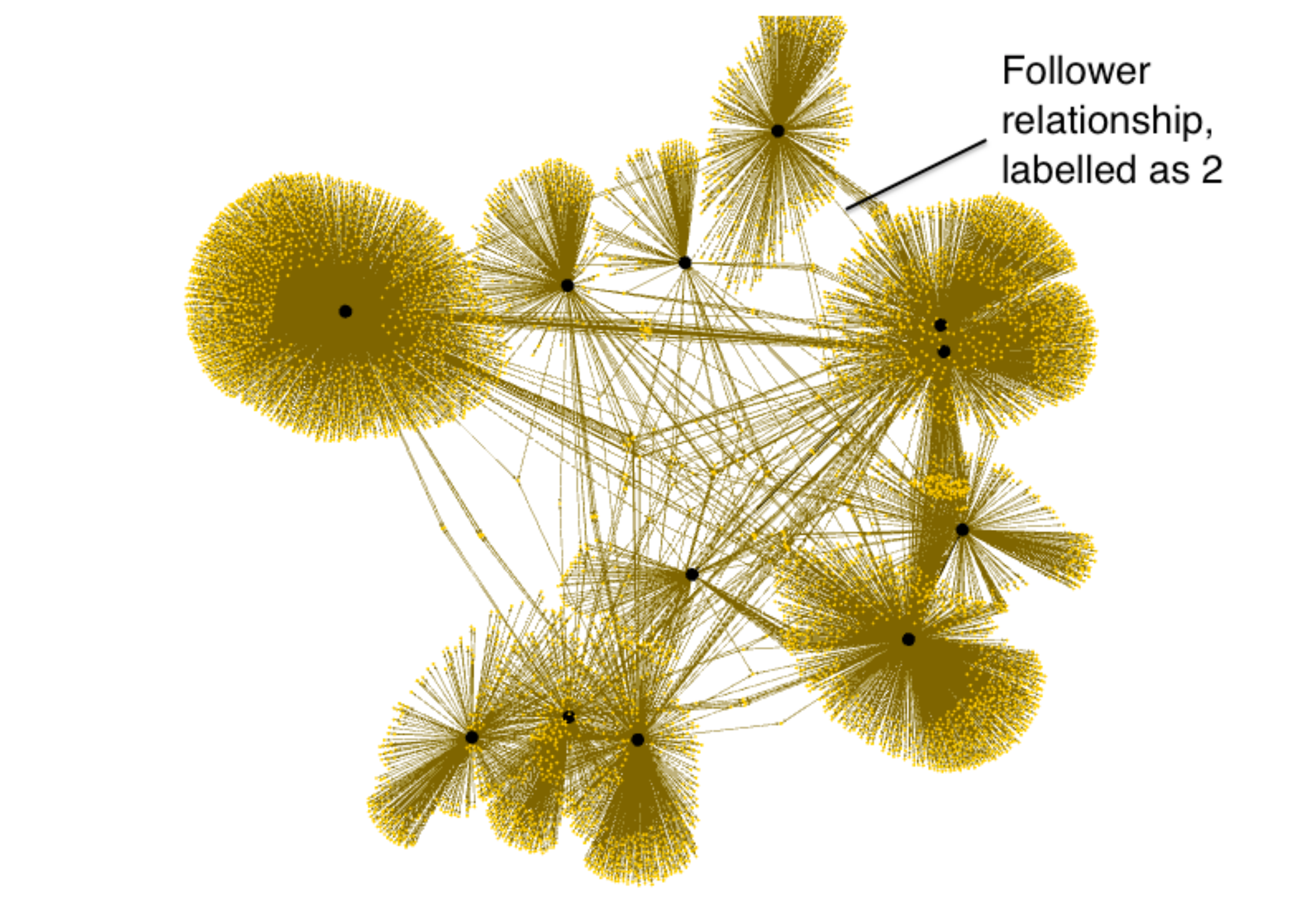}\label{fig:marketing_a}
   }
\quad
   \subfigure[Escort: Audience network]{
   \includegraphics[scale=0.2]{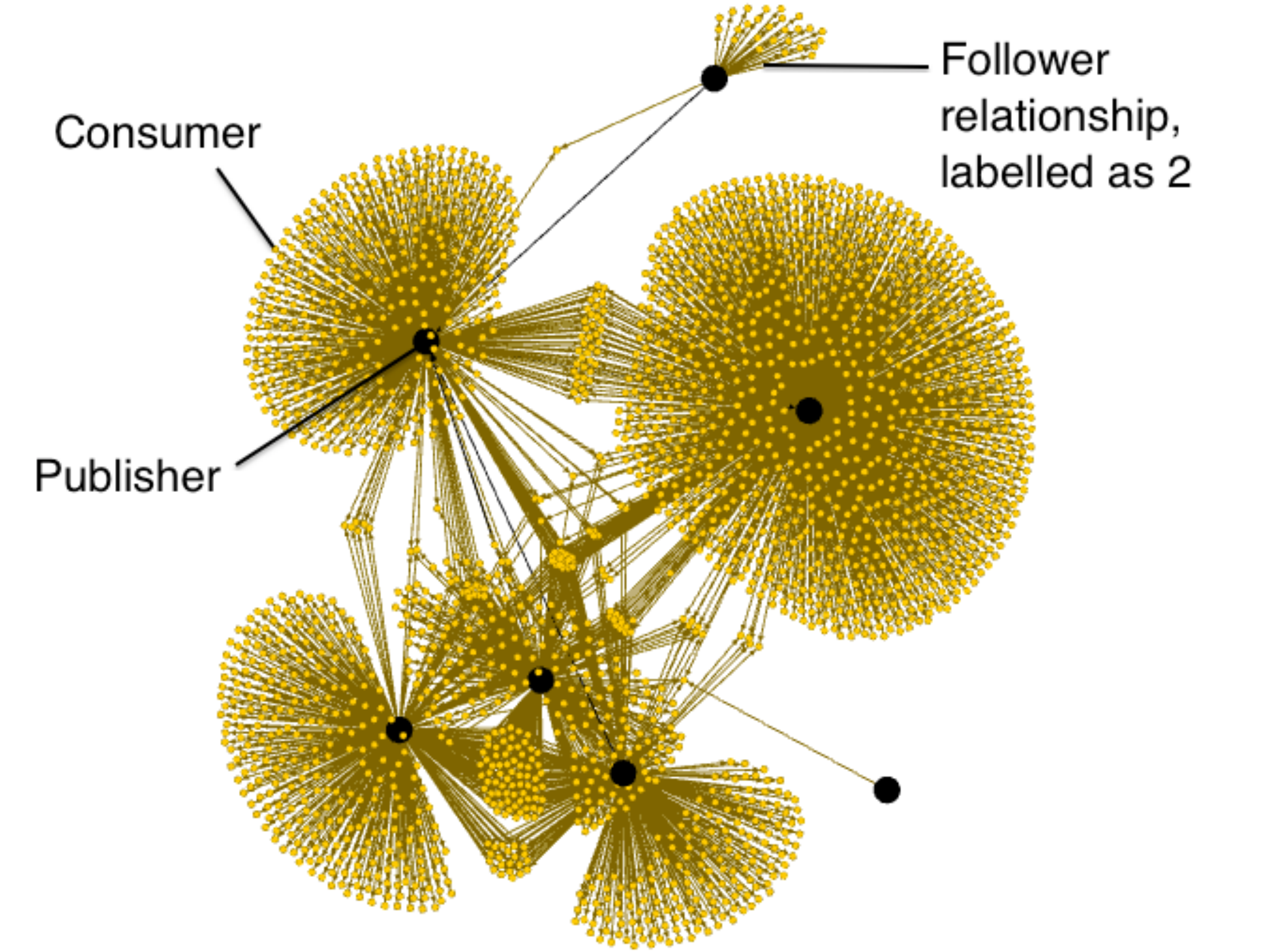}\label{fig:escort_a}
}
\quad
  \subfigure[Emergency: Dissemination network]{
   \includegraphics[scale=0.18]{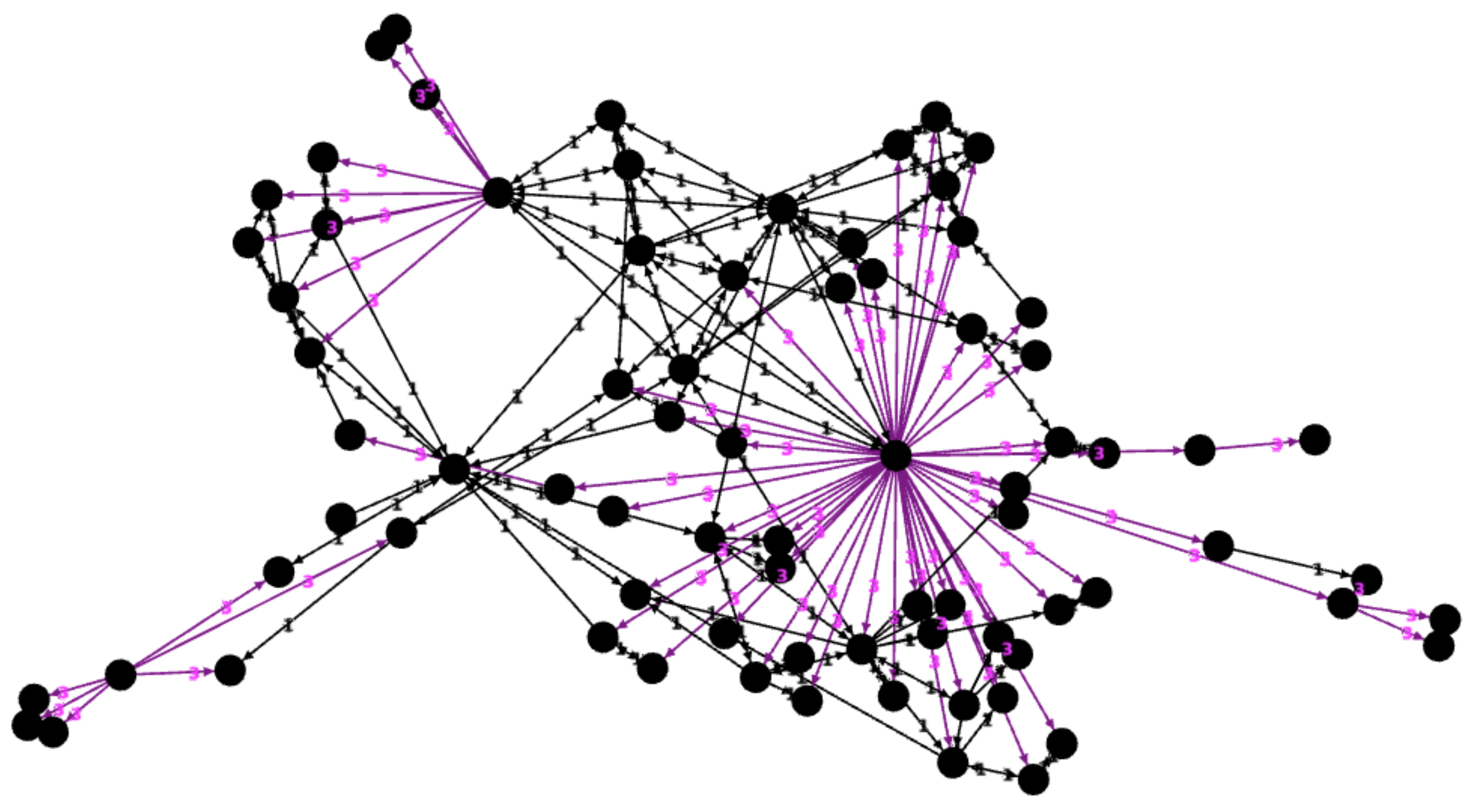}\label{fig:emergency_d}
   }
         \subfigure[\mbox{Entertainment: Dissemination network}]{
   \includegraphics[scale=0.23]{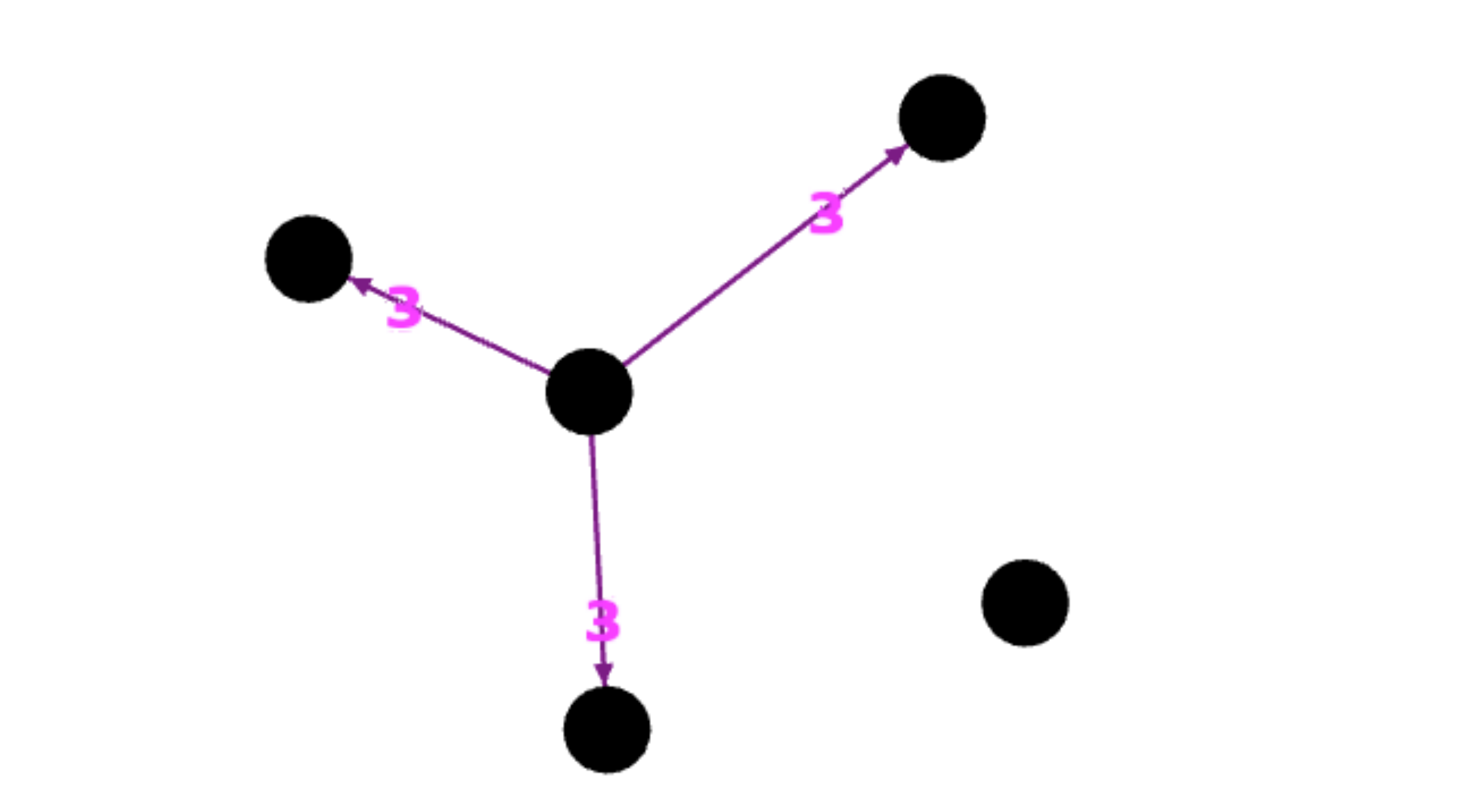}\label{fig:fun_d}
}

\quad
      \subfigure[Entertainment: Dissemination network]{
   \includegraphics[scale=0.2]{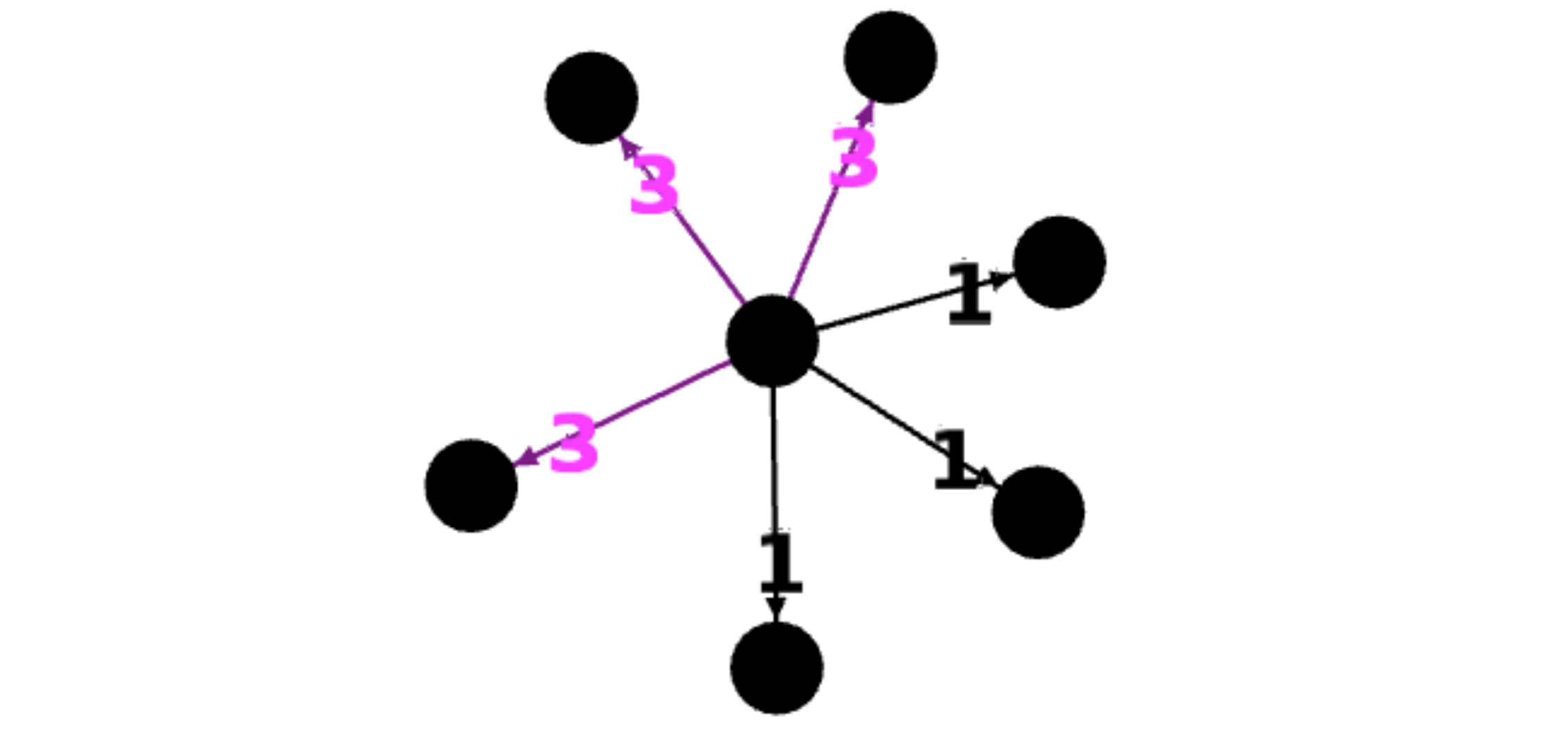}\label{fig:fun_d2}
}
\quad

   \subfigure[\mbox{Marketing: Dissemination network}]{
   \includegraphics[scale=0.28]{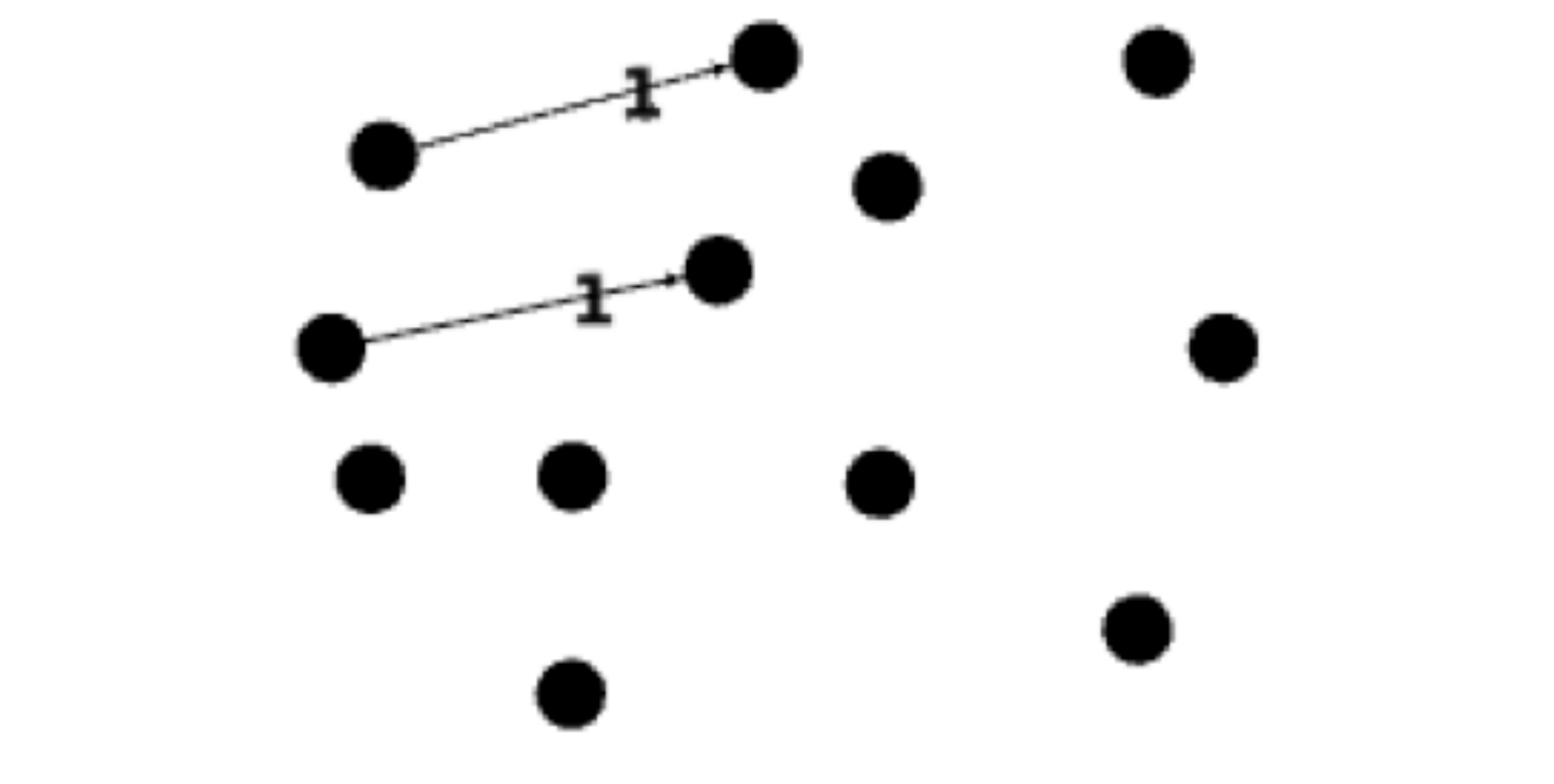}\label{fig:marketing_d}
}
\quad
   \subfigure[{Escort: Dissemination network}]{
   \includegraphics[scale=0.25]{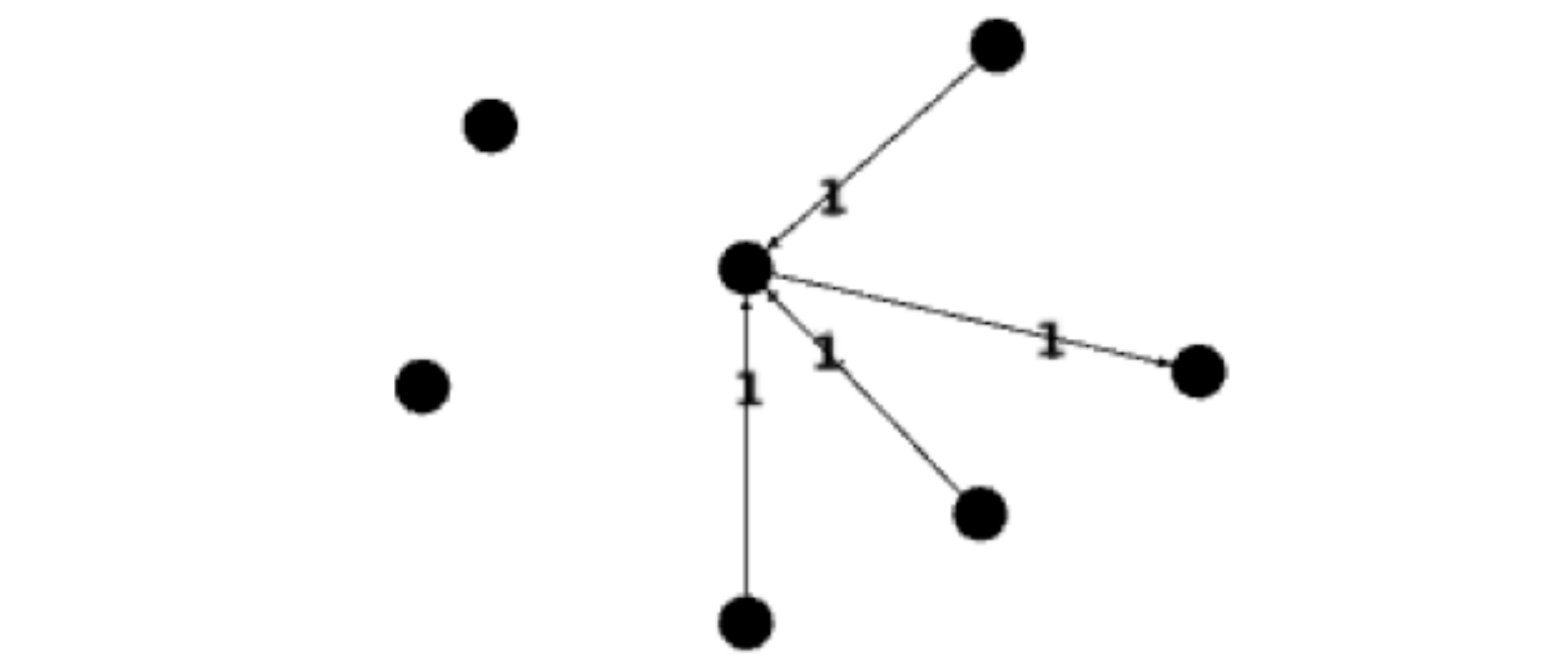}\label{fig:escort_d}
}
   \caption[Audience and dissemination networks formed during most popular mobile number diffusion for each context on Twitter.]{Audience and dissemination networks formed during most popular mobile number diffusion for each context on Twitter -- Emergency, Entertainment, Marketing, Escort business.}
      \label{fig:network}

\end{figure*}

\indent We selected three most popular mobile numbers (in terms of users sharing them, highest being 84 users sharing the same number) in each context, observed in Section \ref{Context}--  emergency, marketing, escort business and entertainment. We explored the audience and dissemination networks for these mobile numbers. We now report the detailed network analysis for emergency context while mentioning briefly the inferences for other contexts. Emergency context includes scenarios where mobile numbers were shared to ask for blood aid, and help. We observed that emergency context was popular on Twitter but not on Facebook. Blood aid was the most popular mobile number shared in the emergency context. The number was posted by few publishers (84) while the information amplified to a wider audience (61,037), fulfilling the intention of popularizing the number.

\noindent \textbf{Audience network}\\
Figure~\ref{fig:emergency_a} shows the
most popular mobile number shared in the emergency context. We observed that audience network was a weakly connected component with high modularity ($>$ 0.8, $>$ 30 communities).
Averaging for the three mobile numbers, average path length of the number numbers in the audience network turns out to be 3, average clustering coefficient was 0.023, and average diameter of the network was 6. We observed that high betweenness nodes, were the publishers in the network. On observation of low clustering coefficient and large path length for the three mobile numbers, we infer that audience networks of a mobile number, shared in the context of an emergency, were loosely bound within one giant component and do not follow small-world phenomena.

\noindent \textbf{Dissemination network}\\
Figure~\ref{fig:emergency_d} shows the dissemination network for the most popular mobile number shared in emergency corresponding to the audience network (Figure~\ref{fig:emergency_a}). Most publishers in the dissemination network were followers (70\%) of other publishers, while few were retweet publishers (30\%). Large proportion of the mobile numbers shared in context of emergency were re-posted and diffused in the network. The average clustering coefficient of the dissemination network was 0.259 and average diameter was 4. Publishers in the dissemination network were closely connected with each other.

We infer that users are sensitive towards emergency scenarios and intend to promote the information by posting the number themselves in their network. For some mobile numbers, we observe that high degree nodes take up such posts (e.g. BloodAid) from normal users who asked for help in emergency, allowing the penetration of information to wider audience. Further, high modularity of audience network ensures that diffused information does not limit to a strongly connected community but diffuse into multiple disjoint communities. Such a network characteristic may be used by government agencies, disaster managers to curb rumors from the network and propagate important information across various communities in the network via active users connected together in a tight-knit component. Note that, the user sharing mobile number has been benefitted by other users' involvement in spreading the information via tight-knit dissemination network.

\begin{table*}[ht]
\begin{center}
\begin{tabular}{|p{2cm}|p{3cm}|p{3cm}|p{3cm}|p{3cm}|}\hline
\textbf{Top numbers} &  \textbf{Emergency} & \textbf{Entertainment} & \textbf{Escort business}  & \textbf{Marketing} \\ \hline
Mobile number 1 & +91-98091-xxxxx & +91-98151-xxxxx  & +91-99002-xxxxx & +91-99002-xxxxx \\ \hline
Mobile number 2 & +91-99110-xxxxx & +91-98551-xxxxx  & +91-99996-xxxxx & +91-96600-xxxxx \\ \hline
Mobile number 3 & +91-99539-xxxxx  & +91-99292-xxxxx & +91-97174-xxxxx & +91-96500-xxxxx \\ \hline
\end{tabular}
\caption{Top mobile numbers extracted for four contexts -- Emergency, Entertainment, Escort Business and Marketing.}
\label{tab:popular_no}
\end{center}
\end{table*}

We conducted similar analysis for other three contexts -- entertainment, escort business and marketing. In entertainment context, we observed only few users (as compared to emergency context) re-posted or re-tweeted the most popular mobile number. Publishers of few popular entertainment mobile numbers were well connected with each other via both follower relationship as well as retweet relationship (see Figure~\ref{fig:fun_d2}), while for other mobile numbers, publishers were majorly connected via retweet relationship (see Figure~\ref{fig:fun_d}). We infer that retweet relationships were promising and were helpful in promotion of entertainment mobile numbers on Twitter, however even the most popular mobile number did not propagate deep in the network. In marketing context, publishers of the mobile numbers were neither connected to each other and nor had overlapping follower networks (see Figure~\ref{fig:marketing_d}). Therefore, marketing numbers disseminate in disjoint networks with few common links among the publishers and fail to diffuse deep in the network. In escort business context, only a few users posted mobile numbers, out of which most users posted it after receiving it on their timeline (i.e. by following original publisher). Less than 10\% of the posts were retweets, implying that users are sensitive on the tweets they should post to their followers (see Figure~\ref{fig:escort_d}). Most popular mobile numbers shared in marketing and escort business context could not receive any intended attention from the social media users.

We conclude that publishers benefit heavily when they share mobile numbers in the emergency context. For other contexts, we do not observe active user participation in disseminating the number and therefore the intent of the publisher to popularize the mobile number is not achieved. We infer that users do not benefit as much as they expose themselves to risks associated with sharing mobile numbers publicly on OSNs.

\chapter{Risk Analysis}\label{chapter:augmentation}
\onehalfspacing
\section{Risk assessment: Risk of collation}
We now turn our focus to understand how publicly shared mobile numbers can be exploited to gather critical and sensitive information about the owners. For this we conducted two experiments.
\subsection{Experiment 1: Collating data from OSNs, Truecaller and Open government data repository (OCEAN)}~\label{experiment}
In this experiment we used two online services -- Truecaller~\footnote{\url{www.truecaller.com}} and home-grown system OCEAN~\cite{Gupta:2013}. Truecaller allows to query a mobile number and returns the name of the owner as well as the network operator. OCEAN, allows to query the name of a person with her finer location in New Delhi (optional) and returns matching entries from publicly available e-government data sources (voter rolls and driving licence records), listing Voter ID, family details, age, home address, and father's name. OCEAN has data only for New Delhi citizens.

\indent We got manual annotators to extract data from Truecaller and OCEAN for Category +91 mobile numbers. For each number, they were asked to observe name of the owner, and her location 
from Truecaller,~\footnote{As per Truecaller policy, we did not store content from it.} along with the name of the owner, and her location from public posts and profiles on OSNs, sharing the same number. Possible names of the mobile number owner and her possible locations were inferred for 2,997 Category +91 numbers. Name of the owners whose inferred location was New Delhi, were then used to query OCEAN and matching set of New Delhi citizens were recorded. Surprisingly, out of annotated 94 New Delhi mobile numbers, we were able to uniquely identify 8 New Delhi users with details like name, age, father's name, home location, gender, and voter ID (see Table~\ref{who:risk}). One of the identified users 
is a professional Indian singer. He posted his number on Facebook and the number revealed other sensitive information about him. We called all the 8 uniquely identified users to validate the information we had about them. Out of 8 people 2 users did not pick our call. From the remaining 6 users 1 user disagreed to answer our questions so we interviewed 5 users. Out of the 5 users only 1 said that the information we had about her was incorrect, however rest confirmed the validity of the information. The incorrect tracing is because the OCEAN system~\cite{Gupta:2013} we used to query user's detail did not have complete data about the citizens of New Delhi. The details of the interview as well as the user's reactions are reported in appendix~\ref{interview}. Aggregation of information extracted from OSNs with the otherwise collected information about a New Delhi mobile number owner, may lead to convenient identity theft attacks~\cite{IDtheft}.

%

\begin{table}[ht]

\begin{center}
\begin{tabular}{|p{9cm}|p{2.4cm}|p{3.3cm}|}
\hline
\raggedright{\textbf{Details}} & \textbf{Shared by owner?} & \textbf{Data source} \\ \hline
\raggedright{+919873xxxxxx, X Kakrania, 24, Male, X Kakrania, ``B-***, B-block, X Vihar Ph-I, Delhi",  WHC17xxx63} & Yes & Facebook, Truecaller, Voter Roll \\ \hline
\raggedright{+9199xxxx2708, X Gambhir, 23, Male, X X Gambhir, ``***, xxxx Bagh, Delhi", NLNxxx5696} & No & Facebook, Truecaller, Voter Roll \\ \hline
\raggedright{+918447xxxxxx, X Singh Nagi, 33, Male, X Singh Nagi, ``D-**-b, Block- D, X Vihar, X Ext., Nangloi",IPN13xxx17} & Yes & Twitter, Truecaller, Voter Roll \\ \hline
\raggedright{+9198xxxx5485, X X Jeswani Pankaj, 53, Male, X X Jeswani, ``***, Mig Flats, *-block, xxxxx Vihar Phase-I", DL/04/xxx/222668 } & Yes  & Facebook, Truecaller, Driving licence records \\ \hline
\end{tabular}
\caption{Anonymized mobile number, name, age, gender, father's name, address, Voter ID of two New Delhi residents who shared their mobile number on OSNs.}
\label{who:risk}
\end{center}
\end{table}

\subsection{Experiment 2: Collating data from OSNs and WhatsApp}
\begin{figure}[!htbp]
  \centering
  \includegraphics[scale=0.5]{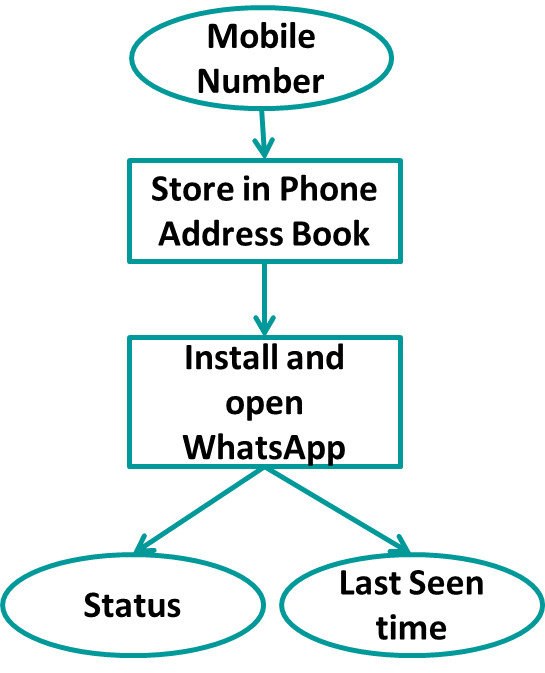}\\
  \caption{Methodology to collect \emph{status message} and \emph{last seen time} of leaked mobile numbers on WhatsApp.}
  \label{fig:whatsApp}
  \vspace{-4mm}
\end{figure}

We experimented with an Android application, WhatsApp,~\footnote{\url{http://www.whatsapp.com/}} to understand if we could infer more sensitive details like \emph{status message} and \emph{last seen time} of the owner of leaked mobile number, and highlight novel significant risks. 
``Last seen time" is the last time when the application user (mobile number owner) went online and accessed the application. To conduct the experiment we used ``Address book matching" feature of WhatsApp and followed the steps shown in Figure ~\ref{fig:whatsApp}.
We added leaked mobile numbers to a phone's contact directory and ran Whatsapp application from the phone \cite{cheng2013bind}. After this we could access 1,071 mobile number (out of 3,076) owner's status message and last seen time on WhatsApp i.e we observed a 34.8\% penetration rate. Penetration rate ($p_{rate}$) is calculated using following formula:
\begin{equation}
p_{rate} = \frac{user_{exposed}}{user_{total}} \times 100
\end{equation}
where $user_{exposed}$ is the number of user accounts exposed on another service and $user_{total}$ is the total mobile numbers tested. Penetration rate indicates the percentage fraction of user accounts exposed on another service (WhatsApp), hence an indicator of disclosure. Mobile numbers which can be connected to a social account brings greater risk to real world identity of its owner. \\
Users leaked variety of sensitive information via their Whatsapp status updates such as travel plans, social network profile, BBM Pins, relationship status. Few examples of status updates are \emph{``100\% Single"}; \emph{``No longer in India. UK: \# +44 75xx 81xxxx US\#610xxxxxxx as of June 10"}; \emph{``www.facebook.com/iakrfi***"}; \emph{``New BBM Pin:  25C7xxxx"}. This information further helps in profiling a user.

\indent We infer that an accidental / unintentional leak of mobile number on OSNs is capable of exposing other sensitive information and thus creating a larger user's digital footprint which may be used against the individual. Beside having a larger footprint, previous literature has shown that the mobile number owner can fall prey to \emph{impersonation attack}, \emph{SMS spam attack}, \emph{Phone number enumeration attack} and \emph{Status message forgery attack} on WhatsApp, if the attacker just have a mobile number in mind (the victim's mobile number) ~\cite{schrittwieser2012guess}.


\section{Risk communication}
With evident risks associated with leaking mobile numbers online, we attempted to communicate the observed risks to mobile number owners. Researchers have suggested various channels for risk communication e.g., Short Message Service (SMS)~\cite{Avivore}, and Interactive Voice Response (IVR) system,~\footnote{\url{http://www.ddm.gov.bd/ivr.php}} to communicate awareness information to its users. Online bloggers have also deployed automated tools to display partially obfuscated mobile numbers onto a public web page~\footnote{\url{http://www.weknowwhatyouredoing.com/}} and SMS with random texts, to publicly shared mobile numbers.~\footnote{\url{http://textastrophe.com/}} We deployed an IVR system and communicated the risks associated with posting mobile number online by calling the owners of the numbers. We chose IVR to ensure the reach to the owners and to convince the credibility of the message to them. We now discuss the IVR deployment details, calling procedure and users' reactions to the calls.


\subsection{IVR system design and implementation}
\begin{figure}[!htbp]
  \centering
  \includegraphics[scale=0.35]{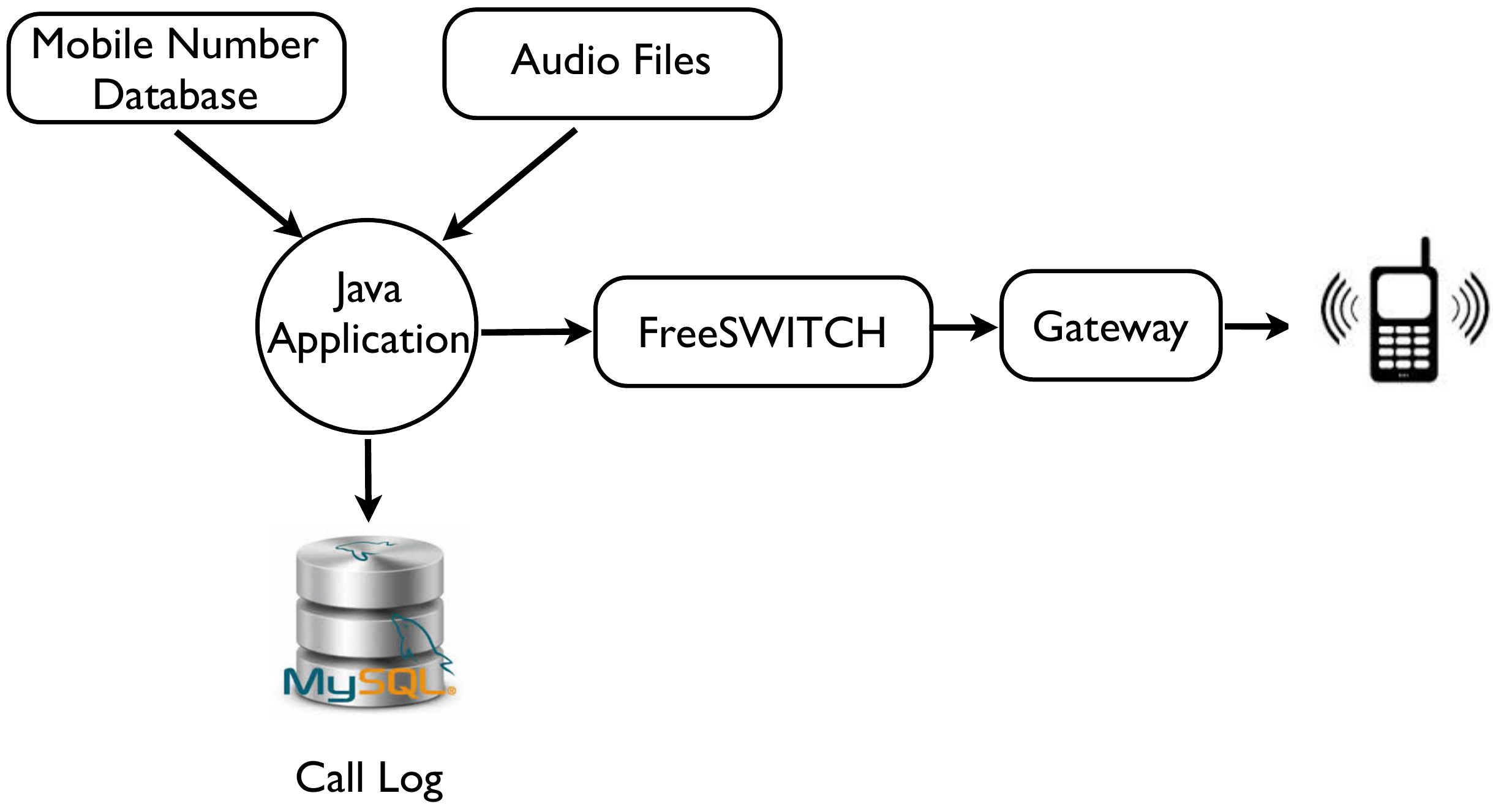}\\
  \caption{IVR System Design implemented using FreeSWITCH and Java application.}
  \label{system_design}
  \vspace{-4mm}
\end{figure}

We setted up an IVR system using FreeSWITCH~\footnote{\url{https://wiki.freeswitch.org/wiki/IVR}} and a Java application (see Figure~\ref{system_design}) to call 2,492 mobile numbers from Category +91, collected from earlier mentioned methodology until 28th February '13.  In India, we were not required to go through an Institutional Review Board (IRB)-type approval process before calling the users. However, 
we applied similar practices in this work. Prior to the actual risk communication part of the message, we informed the user that an audio recording of the call would be taken only for research purposes. Furthermore, participants were given options to disconnect the call and request the deletion of the audio recording, at any given point of time during the call.

\indent When a callee answered the call, for credibility purposes, we introduced ourselves as researchers from New Delhi. We then played the risk communicating message - ``We found your number on X", where X was either ``Facebook" or ``Twitter" or ``Facebook and Twitter", depending on the source from where we extracted the number of the callee. We then prompted a voice message ``Posting your number online is not a good practice. Doing so will make you fall prey to various phone number frauds. Keep yourself safe and consider removing your number from the Internet." We intentionally kept the language simple as English is not a native language of India and we had minimal information about the expertise level of the callee. We then presented callee with the following options: ``Press 1, If you did not know that your number can be leaked and now you will remove it from the Internet; Press 2, If you posted it purposefully and you will not remove it from the Internet; Press 3, if you want to hear the message again." If the user pressed either 1 or 2, we requested him to leave us a feedback and later gave him an option to end the call. We informed the user that we recorded the call for research purposes and logged all responses and activities of callees in a database. We made the calls during weekdays from 1100hrs IST to 1600hrs IST.



\subsection{User reactions}
\begin{figure}[!ht]
  \centering
  \includegraphics[scale=0.6]{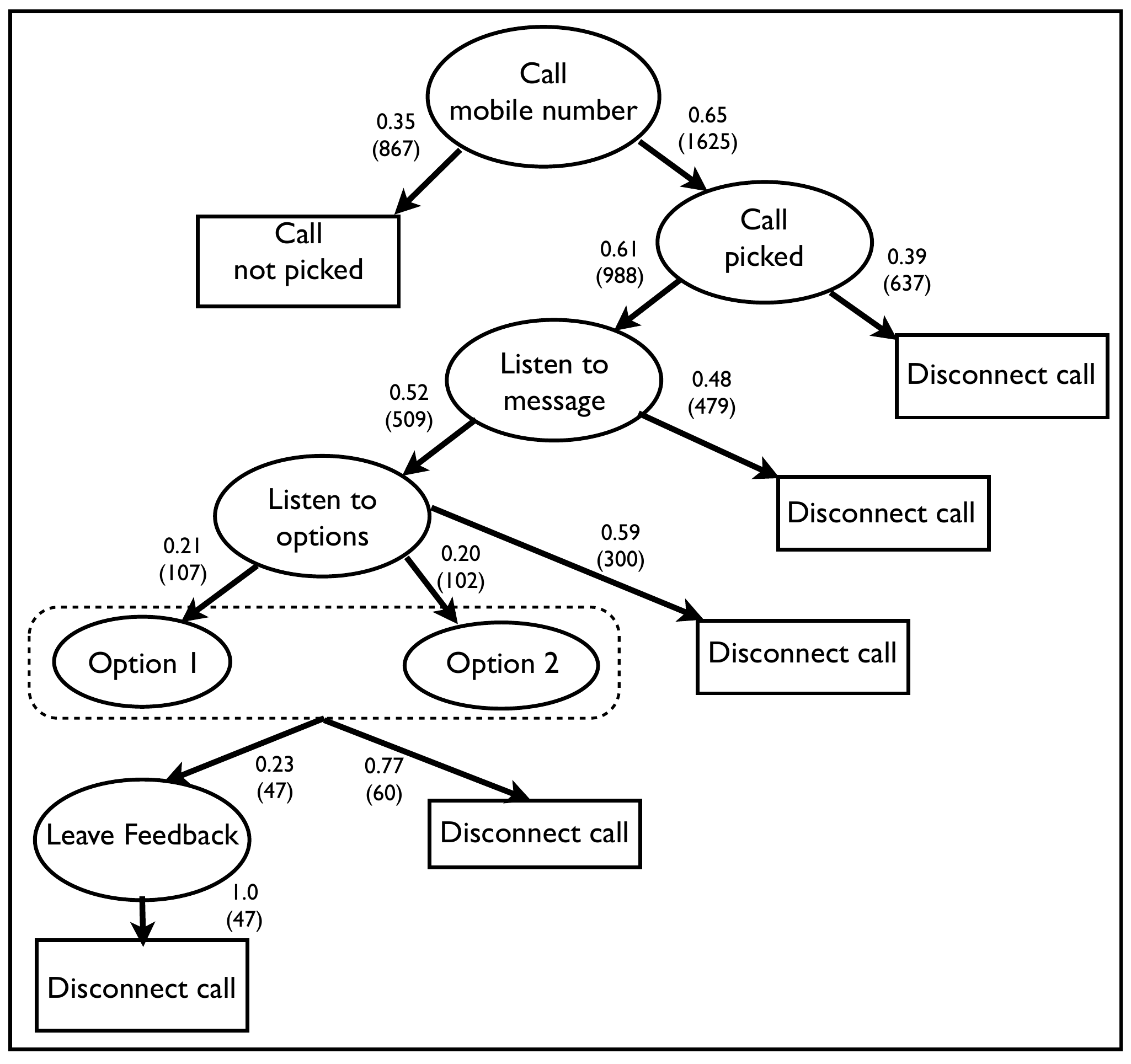}\\
  \caption{Callee Decision Tree. Each stage in the call is associated with a probability and the number of users who chose that stage.}
  \label{callee_decision_tree}
  \vspace{-4mm}
\end{figure}

Figure \ref{callee_decision_tree} shows how callees collectively reacted at each stage during the call. Sixty one percent of callees who picked the call, opted to listen to the message and six percent chose to remove their mobile numbers from OSNs. An equivalent percentage (6.2\%) chose \emph{not} to remove their numbers. Forty seven users from the 2,492 numbers that we called, left feedback on our IVR system.

\subsubsection{User Comments}
We received 47 comments from the mobile number owners, we called. In some of the feedback user's voice wasn't clear or we couldn't hear anything at all. We could understand 26 of them. Following are the broad categories of the comments:

\indent\textbf{``Thank you for information"}\\
All 26 users found the message to be informative and some of them expressed their gratitude to us, for making them aware of the risk of sharing mobile numbers on online social networks. One example feedback was - ``Thank you so much for this very valuable feedback. Bye."

\indent\textbf{``I want to know how to remove my number"}\\
Four callees showed their concern and requested us to help them to remove their numbers from the Internet. One example feedback is - ``I want to know how to remove my number and I don't know, I haven't put my number purposely but if it is there, where exactly it is there I would also like to know that. Please get in touch with me asap. Thank you."

\indent\textbf{``I have deleted, I will not post my number online"}\\
Five callees showed their concerns and proactively removed/promised to remove their mobile number from online social network. One example feedback is - ``Thank you for information, I have deleted, I will not post my number online."\\

Such user reactions urge the necessity for a safeguard solution to control the spread of personal and sensitive information on OSNs.

\indent\textbf{``I posted my number purposely for my website promotion"}\\
Three callees admitted that they shared their number for promoting their business or some other purposes. One example feedback is - ``I posted my number purposely for my website promotion, I usually do deal in web hosting business so that is why I want someone to contact me for hosting services."

Detailed callee reactions/feedback, on receiving the call and listening to the message, spreading awareness about risks of sharing mobile numbers on social media are listed in Appendix Table~\ref{table:User_Feedback}


%
\section{Understanding user's response}
To understand the dynamics of callee's response during the call we looked back at his actual actions (posting mobile number on OSNs). For this we created a set containing all the posts associated with the 107 numbers, whose owners chose option 1, that is they did not know that their number can be leaked and they will remove it from the web. Then we did ownership analysis on the posts present in the set~\ref{section:owner_analysis}. We found 38.3\% (41/107) of mobile numbers were posted by the owner of the number and 0.1\% (9/107) mobile numbers were posted by non-owner of mobile number. Fifty seven tweets during the ownership analysis remained uncategorized.

\indent We observed a disconnect between what users say and what they did. From ownership analysis we found that 38.3\% (41/107) of mobile numbers were posted publicly by their owners, however the owners when contacted said that they did not know that their number can be leaked. A possible reason for this disconnect could be inability of users to manage their privacy settings~\cite{tufekci2008can} or inadvertent disclosure of personal information (mobile number) is also common on online social networks.~\footnote{\url{http://www.dailymail.co.uk/tvshowbiz/article-2072823/Charlie-Sheen-tweets-phone-number-accident-trying-message-Justin-Bieber.html}}

We manually tagged the context of all posts in the set we created. 
We found that 64.5\% (69/107) of mobile numbers were posted for marketing purposes and the remaining 35.5\% were posted in non-marketing context. Marketing context is associated with the post when the intent of the post is to sell or promote something. We called the 8 people whom we uniquely identified to get further insights on the behavior.

\chapter{Discussion, Limitations and Future work}\label{chapter:discussion}
\onehalfspacing
In this work, we examined the exposure of Indian mobile numbers on OSNs via profile and public posts and investigated the associated privacy risks. We collected 76,347 Indian mobile numbers shared on Twitter via tweets or bio and on Facebook via public posts or names and analyzed 2,997 Indian numbers prefixed with +91. Most mobile numbers were shared to ask for blood help, to market astrology business, IT facilities, and escort services. We observed few posts where numbers were shared in personal contexts like \emph{``My contact no in India is +91-9958xxxxxx",} however posts used for personal contexts had few context specific keywords, therefore, personal contexts were difficult to highlight. Males posted their mobile numbers more as compared to females. Users posted same numbers on multiple online social networks and a few of them exploited social aggregators to popularize the number. Though owners promoted mobile numbers via varied methods, most of them did not diffuse deep in the OSN. We observed low benefits but high risks with presence of mobile numbers on OSNs. Sensitive and identifiable information such as Voter ID could be extracted with the use of mobile number and other information sources. However, we acknowledge that such additional information disclosure using a mobile number, is subjective in nature. 
To communicate the risks and vulnerabilities, we called 2,492 numbers with an IVR setup and received feedback. Few users did not know about the presence of their mobile number on OSNs while few told us that they intentionally put it to publicize their business.


\indent Database used in the study might also contribute towards solving \emph{entity resolution problem}, particularly the problem of linking multiple online profiles of a user, to certain extent. But we also suspect that the models leveraging such information might not be generalizable.

\indent Other identifiable information on OSNs such as Blackberry Messenger Pins (BBM), and email addresses, can also help in accurate identification of their owners.~\footnote{We also collected BBM pins and email addresses shared on OSNs.} We infer that though mobile numbers in India are heavily shared for non-personal contexts (e.g. marketing, emergency), such a behavior may invite unwanted spammers / calls to marketers themselves. For instance, Textastrophe~\footnote{\url{http://textastrophe.com/}} pings marketers who post their mobile numbers on public places, hence waste productive hours of the marketer by making inappropriate requests. Textastrophe posts the conversation publicly to demonstrate that leaking mobile numbers may invite unwanted, invalid and untimely requests for marketers too. OSNs do not provide safeguard mechanisms to disallow sensitive and identifiable information exposure via either profile or public posts. There is a need to build technological, people and process oriented solutions to forewarn users and raise the awareness towards risks of mobile number leaks, so that users can make better decisions.

\indent Research communities and industries have been developing technologies and techniques to combat email based phishing attacks. However not much work has been done to prevent \emph{vishing} attacks (voice phishing). Vishing attacks are difficult to detect and stop before they cause harm. Hence it is important to forewarn users and raise awareness amongst them about such attacks. So, in this study we focused on mobile numbers and communicated risks associated to its disclosure to the mobile number owners by setting up an IVR system.

\indent We recognize the limitations of our data collection methodology and analysis. During our keyword selection phase, we used a limited set of keywords to extract posts with mobile numbers from OSNs, and refined keyword set only once. We leave the implications of iterative keyword refinement on the quality of the dataset for future work. During our analysis, we used LIWC tool to tag the words and extract the most frequent categories in which mobile numbers were shared. Unclear and overlapping categories in LIWC output pushed us to manually tag the words, to extract most common contexts in which mobile numbers were shared on OSNs. There is a possibility of self-bias while extracting contexts from words. Our network analysis is limited to Twitter, since Facebook Graph API does not provide public access to users' friends. In future, we plan to analyze Category 0 and Category void numbers to understand their characteristics and examine if the characteristics differ from those of Category +91. We expect future studies to understand impact of geographical, cultural differences, and user personality traits on the practice of sharing mobile number.

\bibliographystyle{plainurl}
\bibliography{references}

\chapter*{Appendix}\label{chapter:appendix}

\section*{Keywords to pull posts having mobile numbers from Twitter and Facebook}
We used Twitter stream API along with a list of keywords to pull public data from Twitter shown in Table \ref{table:Keyword_twitter}.\\
\begin{table*}[!h]
\begin{tabular}{|p{15.5cm}|}
  \hline
{\bf Keywords}\\
  \hline
  \scriptsize{\emph{working phone number}, \emph{my info}, \emph{please call}, \emph{send your number}, \emph{sms me}, mera number, \emph{number}, \emph{call}, \emph{available}, \emph{reachable}, \emph{contact}, \emph{mobile}, \emph{cellphone}, \emph{phone}, \emph{telephone}, \emph{ring}, \emph{cellular}, \emph{directory}, \emph{communicate}, \emph{dial}, \emph{pick}, \emph{beep}, \emph{buzz}, \emph{get back}, \emph{report}, \emph{email}, \emph{phno}, \emph{reach}, \emph{interact}, \emph{connect}, \emph{touch}, \emph{blackberry}, \emph{bbn}, \emph{ph}, \emph{mob}, \emph{num}, \emph{no.}, \emph{whatsapp}, \emph{answer}, \emph{ans}, \emph{text}, \emph{message}, \emph{msg}, \emph{track}, \emph{numbr}, \emph{ph:}, \emph{cell}, RSVP, lost my number, \emph{phnum}, \emph{reach}, \emph{interact}, \emph{check with}, get free gifts $+91$, get money, free cards call now, call now, sell, carpool contact, carpool call, carpool mumbar call, carpool sms, sexy $+$91, lost, mms $+91$, sms $+91$, sale $+91$, buy $+91$, discount $+91$, information $+91$, festival $+91$, competition $+91$, college event $+91$, my number is $+91$, call me at, contact $+91$, proposal and presentation $+91$, numerologist $+91$, hotel $+91$, photographer $+91$, exhibition $+91$, free dicount $+91$, studio $+91$, advocate $+91$, advocate 0, astrologer 0, artist $+91$, artist 0, insurance 91, insurance 0, fashion 0, radio 0, fashion $+91$, radio $+91$, astrologer $+91$, call @ $+91$, lost my number, my phone number, text me, $+91$, my new number, girls number, mob:, number:, call this number, cell number $+91$, sell number $+91$, lost all contacts, sexy boys number, buy number, GSM number $+91$, emergency call $+91$, number $+$91 free gifts, discount number, sale phone number, sexy girls number, check with, ring on 91, call me 91, contact me 91, calling card 91, contact $+91$, urgently $+91$, blood aid $+91$, bloodaid $+91$, blood $+91$, register $+91$, helpdesk $+91$, urgeny $+91$, escort $+91$, male $+91$, house $+91$, rent $+91$, molvi $+91$, magic $+91$, relief $+91$, project $+91$, love $+91$, event $+91$, company $+91$, prayers $+91$, plz $+91$, please $+91$, interview $+91$, street $+91$, offer $+91$, christmas $+91$, carpool contact $+91$, carpool sms $+91$, sexy $+91$, lost $+91$, my number is $+91$, call me at $+91$, proposal and presentation $+91$, numerologist $+91$, hotel $+91$, photographer $+91$, exhibition $+91$, studio $+91$, advocate $+91$, astrologer $+91$, artist $+91$, insurance $+91$, fashion $+91$, radio $+91$, astrologer $+91$, call @ $+91$, contact $+91$, lost my number $+91$, my phone number $+91$, text me $+91$, $+91$, my new number $+91$, girls number $+91$, mob: $+91$, number: $+91$, call this number $+91$, cell number $+91$, sell number $+91$, lost all contacts $+91$, sexy boys number $+91$, buy number $+91$, GSM number $+91$, emergency call $+91$, free gifts $+91$, work phone number $+91$, my info $+91$, please call $+91$, discount number $+91$, sale phone number $+91$, sexy girls number $+91$, send your number $+91$, sms me $+91$, mera number $+91$, number $+91$, available $+91$, reachable $+91$, mobile $+91$, cellphone $+91$, phone $+91$, telephone $+91$, cellular $+91$, directory $+91$, communicate $+91$, dial $+91$, pick $+91$, beep $+91$, buzz $+91$, get back $+91$, report $+91$, reach $+91$, interact $+91$, ring on $+91$, call me $+91$, contact me $+91$, calling card $+91$, contact $+91$, sevice $+91$, hot girls $+91$, connect $+91$, touch $+91$, blackberry $+91$, bbn $+91$, ph $+91$, mob $+91$, num $+91$, no. $+91$, whatsapp $+91$, answer $+91$, ans $+91$, text $+91$, message $+91$, msg $+91$, track $+91$, numbr $+91$, ph: $+91$, cell $+91$, RSVP $+91$, lost my number $+91$, ring $+91$, cellular $+91$, directory $+91$, communicate $+91$, dial $+91$, pick $+91$, beep $+91$, buzz $+91$, get back $+91$, report $+91$, email $+91$, phno $+91$, phnum $+91$, reach $+91$, interact $+91$, check with $+91$, sale $+91$, buy $+91$, discount $+91$, information $+91$, guru $+91$, vashikaran $+91$, sevice $+91$, festival $+91$, competition $+91$, college event $+91$} \\ \hline

\end{tabular}
\caption{Keywords used to pull data from Twitters Stream API and Facebook Graph API. Italicized words were initial 50 keywords.}
\label{table:Keyword_twitter}
\end{table*}

\pagebreak

\section*{User feedback}

Users left us the following feedbacks:

\begin{table*}[!h]
\begin{tabular}{|p{15.5cm}|}
\hline

\scriptsize{I wasn't aware about the activity as I posted, so did my friends. I didn't know about the side effects of the number. Thank you}\\
\hline
\scriptsize{thanks}\\
\hline
\scriptsize{It is a very nice process that you are doing and making people aware about online frauds and telephone number frauds but your system is basically selling business houses where phone numbers need to be present online, so you need to improve upon your system}\\
\hline
\scriptsize{Yeah thank you}\\
\hline
\scriptsize{Thanks for informing}\\
\hline
\scriptsize{Thank you so much for this very valuable feedback. Bye}\\
\hline
\scriptsize{Thank so much for the information, I will remove it from a Facebook account. Thank you so much}\\
\hline
\scriptsize{Please guide me what to do. Actually, I don't know what is the procedure and what is the system}\\
\hline
\scriptsize{Hello already posted my number for gmail columns, I will remove it from the internet. Thank you}\\
\hline
\scriptsize{I am not getting your point properly}\\
\hline
\scriptsize{Hello *******}\\
\hline
\scriptsize{I didn't know that my number is there on Facebook. Kindly delete it}\\
\hline
\scriptsize{I will delete it}\\
\hline
\scriptsize{Mam thank you for information, I have deleted, I will not post my number online. Thank you for the information}\\
\hline
\scriptsize{Please remove this * my number}\\
\hline
\scriptsize{Yeah hi got a call from Precog from * I didn't get anything could you make me a call and give me a proper details. Thank you}\\
\hline
\scriptsize{Yes I have posted my mobile no for some kind of purpose. Thanks}\\
\hline
\scriptsize{I want to how to remove my number and I don't know, I haven't put my number purposely but if it is there where exactly it is there I would also like to know that. Please get in touch with me asap. Thank you}\\
\hline
\scriptsize{I dont want my number to be post in Facebook, yeah}\\
\hline
\scriptsize{Hello I have posted my number on Facebook * I will be getting the Facebook updates * on the mobile number}\\
\hline
\scriptsize{I registered my cell number on the internet forum * purpose of business and thanks for remember remind me and aware me about the disadvantage of the registration of this cell number on internet but I not aware, I want to be registered my number, I want to be continued on internet due to my business purpose and thanks for informing. In fact i didn't know that number shouldn't be given. I am sorry for that. Thank you}\\
\hline
\scriptsize{I posted my no purposely for my website promotion, I usually do deal in web hosting business so i like,thats a I want someone to contact me for like hosting services and all}\\
\hline
\scriptsize{Hello my name is *** and my number is *** I have purposely given my number in Facebook, it is because of the reason that I am running a business and my business involves with some marketing strategies as well for which I want clients to give me a feedback or either call me up regarding works that I did with them, So I have purposely given this and I wont like to remove my number. Thank you}\\
\hline
\scriptsize{Hi, I am in a, I have a very small company. The name of the company  silver * . So I have posted this number on Facebook. Hello}\\
\hline
\scriptsize{I will do this *** from the Twitter}\\
\hline
\scriptsize{Thank for your suggestion}\\
\hline
\end{tabular}
\caption{User Feedback}
\label{table:User_Feedback}
\end{table*}

\pagebreak

\section*{Interview Details}~\label{interview}

We called all the 8 subjects whom we were able to uniquely identify~\ref{experiment} to validate the information we had about them. When a callee answered the call, for credibility purposes, we introduced ourselves as researchers from Precog at IIIT-Delhi. We then told them -- ``We found your number on X", where X was either ``Facebook" or ``Twitter" or ``Facebook and Twitter", depending on the source from where we extracted the number of the callee. We then requested them to help us in our research by saying -- ``We will be grateful if you can help us in our research. We assure you that we will not use your details for any unauthorized purpose". Then we asked questions confirming their personal details we had about them. After this we enquired if they posted their number purposefully on OSNs and why. We also asked if they plan to remove their number from the internet and why. After this we thanked them for their responses and communicated how risky can it be when they share their number on OSN by saying - ``We inferred all the details from your public social media profile and public e-government sources. We hope that now you understand that posting your number online is not a good practice. Doing so might make you fall prey to various phone number frauds. Keep yourself safe and consider removing your number from the internet". Later we asked if they have any feed back for us and ended the call after taking the feedback and answering all the questions, if they had any.

\indent Out of 8 people 2 users did not pick our call. Out of the remaining 6 users 1 user disagreed to answer our questions so we interviewed 5 users. From the 5 users only 1 said that the information we had about her was incorrect, however rest confirmed the validity of the information. The incorrect tracing is because the OCEAN system~\cite{Gupta:2013} we used to query user's detail did not have complete data about the citizens of New Delhi.

\indent Of the users we interviewed, two users expressed concern and said they only wanted to share their number with friends and relatives. On knowing about the possible leakage and associated risks, they decided to remove their number. One user also went ahead and called his service provider and confirmed that we can very easily block his number and do much more with the information we had. The user said that he shared his number so that his customers can reach him, however he understood the risks associated to it and requested for possible countermeasures in such a situation. He also suggested that the mobile service providers can use a stronger authentication technique like asking questions on last recharge details or last service detail from the customer.

\end{document}